\documentclass{emulateapj}
\usepackage{txfonts}
\usepackage[utf8x]{inputenc}
\usepackage{graphicx}
\usepackage{natbib}
\usepackage{url}
\newcommand{\nuLnu}{$\nu$L$_\nu$}

\newcommand{\Lsun}{L$_\odot$}

\slugcomment{Draft \today}
\shorttitle{Evolving infrared SEDs and star formation rate estimates}
\shortauthors{Nordon et al.}
\begin{document}
\title{The impact of evolving infrared spectral
energy distributions of galaxies on star formation rate estimates}

\author{
R.~Nordon\altaffilmark{1}, 
D.~Lutz\altaffilmark{1},
R.~Genzel\altaffilmark{1},
S.~Berta\altaffilmark{1}, 
S.~Wuyts\altaffilmark{1}, 
B.~Magnelli\altaffilmark{1},
B.~Altieri\altaffilmark{2},
P.~Andreani\altaffilmark{3},
H.~Aussel\altaffilmark{4},
A.~Bongiovanni\altaffilmark{5,6},
J.~Cepa\altaffilmark{5,6},
A.~Cimatti\altaffilmark{7},
E.~Daddi\altaffilmark{4},
D.~Fadda\altaffilmark{8}, 
N. M.~F{\"o}rster Schreiber\altaffilmark{1}, 
G.~Lagache\altaffilmark{9},
R.~Maiolino\altaffilmark{10},
A. M.~P{\'e}rez Garc{\'i}a\altaffilmark{5,6},
A.~Poglitsch\altaffilmark{1}, 
P.~Popesso\altaffilmark{1},
F.~Pozzi\altaffilmark{10},
G.~Rodighiero\altaffilmark{11},
D.~Rosario\altaffilmark{1}, 
A.~Saintonge\altaffilmark{1},
M.~Sanchez-Portal\altaffilmark{2},
P.~Santini\altaffilmark{1,10}, 
E.~Sturm\altaffilmark{1}, 
L.J.~Tacconi\altaffilmark{1},
I.~Valtchanov\altaffilmark{2},
L.~Yan\altaffilmark{8},
}
\altaffiltext{1}{Max-Planck-Institut f\"ur extraterrestrische Physik,
Postfach 1312, 85741 Garching, Germany \email{nordon@mpe.mpg.de}}
\altaffiltext{2}{Herschel Science Centre, European Space Astronomy Centre, ESA, Villanueva de la Ca\~{n}ada, 28691 Madrid, Spain}
\altaffiltext{3}{ESO, Karl-Schwarzschild-Str. 2, D-85748 Garching, Germany}
\altaffiltext{4}{Laboratoire AIM, CEA/DSM-CNRS-Universit{\'e} Paris Diderot, IRFU/Service d'Astrophysique,
B\^at.709, CEA-Saclay, 91191 Gif-sur-Yvette Cedex, France}
\altaffiltext{5}{Instituto de Astrof{\'i}sica de Canarias, 38205 La Laguna, Spain}
\altaffiltext{6}{Departamento de Astrof{\'i}sica, Universidad de La Laguna, Spain}
\altaffiltext{7}{Dipartimento di Astronomia, Universit{\`a} di Bologna, Via Ranzani 1,
40127 Bologna, Italy}
\altaffiltext{8}{IPAC, California Institute of Technology, Pasadena, 
CA 91125, USA}
\altaffiltext{9}{Institut d'Astrophysique Spatiale (IAS), B\^at 121, Universit\'{e} de Paris XI, 91450 Orsay Cedex, France}
\altaffiltext{10}{INAF - Osservatorio Astronomico di Roma, via di Frascati 33,
00040 Monte Porzio Catone, Italy}
\altaffiltext{11}{Department of Astronomy, University of Padova, Vicolo dell'Osservatorio 3, 35122, Padova, Italy}

\begin{abstract}
We combine {\it Herschel}-PACS\footnote{Herschel is an ESA space observatory 
with science instruments provided by European-led Principal
Investigator consortia and with important participation from NASA.} 
data from the PEP program with {\it Spitzer} 24~$\mu$m and 16~$\mu$m 
photometry and ultra deep IRS mid-infrared spectra to measure 
the mid- to far-infrared spectral energy distribution (SED) of $0.7<z<2.5$
normal star forming galaxies around the main sequence
(the redshift-dependent relation of star formation rate and stellar mass).
Our very deep data confirm from individual far-infrared detections
that z$\sim$2 star formation rates are overestimated if based on 24~$\mu$m 
fluxes and SED templates that are calibrated via local trends with luminosity.
Galaxies with similar ratios of rest-frame \nuLnu(8) to 8-1000~$\mu$m infrared 
luminosity (LIR) tend to lie along lines of constant offset from the 
main sequence.
We explore the relation between SED shape and offset in specific 
star formation rate (SSFR) from the redshift-dependent main sequence.
Main sequence galaxies tend to have a similar \nuLnu(8)/LIR regardless of 
LIR and redshift, up to z$\sim$2.5, and \nuLnu(8)/LIR decreases 
with increasing offset above the main sequence in a consistent way at the 
studied redshifts. 
We provide a redshift-independent calibration of SED templates in the 
range of 8--60 $\mu$m as a function of $\Delta$log(SSFR) offset 
from the main sequence. Redshift dependency enters only through the 
evolution of the main sequence with time.
Ultra deep IRS spectra match these SED trends well and verify that they
are mostly due to a change in ratio of PAH to LIR rather than continua
of hidden AGN.
Alternatively, we discuss the dependence of \nuLnu(8)/LIR on LIR.
Same \nuLnu(8)/LIR is reached at increasingly higher LIR at higher redshift, 
with shifts relative to local by 0.5~and 0.8~dex in log(LIR) at redshifts 
z$\sim$1 and z$\sim$2. Corresponding SED template calibrations are provided 
for use if no stellar masses are in hand.
For most of those z$\sim$2 star forming galaxies that also host 
an AGN, the mid-infrared is dominated by the star forming component.
\end{abstract}


\maketitle

\section{Introduction}
In any study of galaxy evolution, the star formation rate (SFR) is a key 
parameter. SFRs can be estimated in various ways from different 
wavelengths, but for all reasonably massive and dusty star forming galaxies,
the far-infrared emission will dominate the emitted power and provide 
a reliable SFR estimator.
In star forming galaxies (SFG) most of the light which is emitted from 
young stars is absorbed by dust and re-emitted in the far-infrared, peaking 
around 60--100~$\mu$m rest wavelength.
Thus, a measurement or an estimate of the total infrared luminosity (LIR, 
integrated between rest frame 8 and 1000 $\mu$m) of a galaxy is required in 
order to account for the reprocessed fraction of the light.

The inaccessibility of the far-infrared to ground based instruments and the 
technological challenges of far-infrared space missions have, until recently, 
left this critical part of the spectrum largely unexplored in high redshift 
galaxies. Other wavelengths have been used instead to estimate LIR and the 
mid-infrared, in particular observed by the {\it Spitzer} space telescope, 
has been important in detecting high redshift, star forming and dust obscured 
galaxies. The mid-infrared itself however accounts for only a small fraction 
of the LIR and at increasing redshifts, the mid-infrared band covers even 
shorter rest-frame wavelengths, on the edge of the relevant 8--1000 $\mu$m 
range. The derivation of LIR from mid-infrared observations requires large 
extrapolations that carry significant uncertainties. These extrapolations 
are typically based on families of SEDs that are parametrized as a function 
of LIR and calibrated to match the trends observed in the local universe, in
particular the library of \citet[][CE01 in the following]{CE01}.  

Indications that local infrared SEDs may not hold emerged with 
the first more detailed high-z studies. Luminous sub millimeter galaxies (SMGs) were
inferred to have lower far-infrared dust temperatures  
\citep[e.g.,][]{Chapman05, Pope06}, and larger Polycyclic Aromatic Hydrocarbon
(PAH) equivalent width and ratio to LIR \citep[e.g.,][]{Lutz05, Valiante07, Menendez07, Pope08, Murphy09}, when compared to local 
sources of equivalent LIR. Strong PAH emission was also found in some
luminous `dust obscured galaxies' (DOGs) selected by high 24~$\mu$m to optical
flux ratios \citep[e.g.,][]{Weedman06, Sajina07b, Murphy09}. Both types of 
selections induce SED bias and studies typically referred to luminous 
LIR$\gtrsim$10$^{12.5}$~L$_\odot$ sources. These limitations much less apply 
to studies of a small number of lensed galaxies
\citep{Desai06, Rigby08, Siana08, Siana09, Fadely10}, again typically 
indicating changing SEDs and stronger PAH than in local galaxies of similar LIR. 
Using 70 and 160~$\mu$m stacking of 24~$\mu$m sources \citep{Papovich07},
radio stacking of BzK galaxies \citep{Daddi07a} and an SED study including
IRS spectra of a sample of SMGs, DOGs, and AGN hosts \citep{Murphy09},
evidence was then found that extrapolations from Spitzer 24~$\mu$m 
mid-infrared photometry
tends to overestimate LIR for z$\sim$2 galaxies by factors of 2--10.
This concerns a key epoch of galaxy evolution, during which both cosmic star 
formation and AGN activity peak, and which is thus at the focus of many 
current observational and theoretical studies.
Whether this relative `mid-infrared excess' is due to continuum emission 
from dust heated by obscured AGNs that boost mid-infrared fluxes 
\citep{Daddi07b}, or due to a mismatch between the real SEDs at high-z and the locally-calibrated 
templates that are used in converting the mid-infrared flux to LIR, remained 
unclear. Stacking of X-ray data revealed a large fraction of 
obscured AGN in `mid-infrared excess' sources \citep{Daddi07b},
but then evidence from IRS spectroscopy pointed towards strong emission from 
 (PAH) not fully accounted for by the templates
\citep{Murphy09, Fadda10}.

At z$>$1.5 the strong rest-frame 8~$\mu$m PAH complex is redshifted into
the 24~$\mu$m filter, which raised the suspicion that the local template SEDs 
have too low ratios of PAH to far-infrared emission, which in turn leads to 
inaccurate extrapolation. This ratio varies 
significantly among local galaxies. Extrapolations from 8~$\mu$m flux to 
LIR thus critically depend on selecting the correct SED template.

The 8~$\mu$m PAH complex includes several broad emission bands. Its flux 
relative to the far-infrared 
depends strongly on the local ISM conditions 
\citep{Laurent00,Sales10}. In the local SED family of CE01, the ratio of
8$\mu$m PAH to LIR is fairly constant at LIR$\lesssim$10$^{10}$~L$_\odot$ 
where it is based mostly on normal star forming disks. At higher LIR where the
ratio  of the 8~$\mu$m luminosity to LIR decreases, it is 
based on low-redshift luminous infrared galaxies 
(LIRG, LIR$>$10$^{11}$~L$_\odot$) and ultra-luminous infrared galaxies 
(ULIRG, LIR$>$10$^{12}$~L$_\odot$). Local (U)LIRGs have distinct properties
that make them differ from an aggregation of numerous
disk star forming regions.
They are typically interacting or merging systems
\citep{Sanders96} with their luminous star formation concentrated in a small 
region of size few 100~pc \citep{Condon91}. They show high 
`star formation efficiencies' SFR/M$_{\rm Gas}$, lower CO luminosity to gas mass 
conversion factor \citep{Solomon97, Downes98}, and a `deficit' of [CII] 158 
compared to LIR \citep[e.g.,][]{Malhotra97, Luhman03, Gracia11}.
All these properties reflect a compact structure and intense radiation fields.
The intense radiation fields naturally lead to
lower PAH/LIR and a warmer far-infrared dust peak \citep[e.g.][]{DH01}, placing
ULIRGs at the high temperature end of the local compactness-temperature
relation \citep{Chanial07}.

All these well established properties and connotations of the local 
ULIRGs, beyond the basic definition of the class by LIR, do not necessarily 
apply for equally luminous z$\sim$2 galaxies.
High redshift galaxies with ULIRG-like luminosities are more 
gas rich \citep[e.g.,][]{Tacconi10}, can be disk-like galaxies without 
indication of recent mergers 
\citep[e.g.,][]{Shapiro08, ForsterSchreiber09} and are not as compact as the 
local ULIRGs \citep{Bouche07, Kriek09, ForsterSchreiber11a}.
Along with this come arguments for more normal star formation 
efficiencies \citep{Genzel10, Daddi10b} and less of a [CII] deficit 
\citep[e.g.,][]{Maiolino09, Gracia11}. Changes in
star formation efficiencies and [CII] deficit reflect changing
local ISM conditions in high-redshift galaxies compared to local ones of 
the same luminosity, and challenge the association of the absolute 
LIR with a single template SED shape.

In recent years, a correlation has been established between the 
star formation rate (SFR) and stellar mass of star forming galaxies 
\citep{Noeske07,Daddi07a,Elbaz07}. This is now commonly referred to as 
the `main sequence' of star forming galaxies, from which passive early type 
galaxies and local ULIRGs are outliers. 
The absolute scaling and possibly the slope of the specific star formation 
rate (SSFR, i.e. star formation rate per unit stellar mass) main sequence evolve with redshift. The SSFR significantly increases with redshift, scaling as 
$(1+z)^n$, where $2.2<n<4$ \citep{Daddi07a, Erb06, Damen09, Dunne09, 
Pannella09, Bouche10, Rodighiero10, Karim11}. There is an ongoing debate on 
the slope
$\beta$ of the SSFR$\propto$M$_*^\beta$ main sequence of star forming galaxies. At 
z$\sim$1--2, much of the variation that is found may be related to how 
specifically
star forming galaxies are separated from passive ones. BzK color selections
seem to produce flatter slopes near $\beta=-0.1$ \citep{Daddi07a, Pannella09} 
compared to $\beta=-0.4\ldots-0.5$ when isolating SFGs in other ways from 
original mass selections \citep{Rodighiero10, Karim11}, see also the direct 
comparison in \citet{Karim11}.

One of the implications of this fairly tight correlation between SFR and mass
is that star formation in these objects must be relatively continuous. This 
raises yet another reason to question the applicability of mid- to 
far-infrared flux ratios that were calibrated on local ULIRGs 
with strongly peaked star formation histories. The possibility arises that 
SSFR or its offset from the main sequence may more generally define the SED 
shape, rather than LIR alone.

With the advent of the {\it Herschel} space telescope \citep{Pilbratt2010} 
and its PACS instrument \citep{Poglitsch2010}, main sequence SFGs at 
redshifts of z$\sim$1--2  can now be detected 
directly in the 100 and 160~$\mu$m bands, reaching objects with star 
formation rates as low as 10s to a few 100s M$_\odot$~yr$^{-1}$ and typical 
masses above 10$^{10}$~M$_\odot$. 
This is due to the Herschel surveys reaching down to very faint fluxes of 
order 1~mJy \citep{Lutz11, Berta11}. 
New Herschel-PACS results confirmed and extended the finding that 
extrapolations from observed 
24~$\mu$m wavelengths for z$>$1.5 galaxies tend to overestimate LIR by 
factors of 4 to 7 \citep{Nordon10, Elbaz10}. This is in contrast to lower 
redshifts where the extrapolations are reasonably accurate \citep{Elbaz10}.
Using data from the GOODS-Herschel program, \citet{Elbaz11}
study the ratio of total IR luminosity to rest-frame 8 $\mu$m luminosity,
which they find to be universal for most star forming galaxies, defining an IR main sequence
in SFR vs. mass.
The outliers (enhanced SFR) to this relation are found to be starbursting galaxies with 
high projected star formation surface densities.
We compare our work to their results in the discussion.

In this paper we study the relations between the mass, star formation rate 
(also LIR) and mid- to far-infrared SED properties of distant star forming 
galaxies.
We focus on two redshift bins: $1.5<z<2.5$ and $0.7<z<1.3$ where the 8~$\mu$m 
PAH emission complex is covered by the {\it Spitzer} MIPS 24~$\mu$m and 
IRS blue peak-up (16~$\mu$m) filters (MIPS24 and IRS16 hereafter) respectively. 
{\it Herschel}-PACS provides the crucial LIR measurements and samples the 
far-infrared SED shape. We find that galaxies with constant \nuLnu(8~$\mu$m) 
(where \nuLnu$=4\pi D_{lumin}^2 \nu F_\nu$)
to LIR emission ratios tend to lie in lines parallel to the main sequence 
(Section~\ref{sec:iso nuLnu8/LIR}). This motivates us to study 
the SED properties of galaxies as a function of their distance from the 
redshift-dependent main sequence, in addition to the traditional view as a 
function of IR luminosity.
We adopt the redshift dependent main sequence from \citet{Rodighiero10}, which
is based on {\it Herschel}-PACS data closely related to our sample.

In Sections~\ref{sec:Data} and  \ref{sec:Excess} we describe our sample, 
reconfirm from our deeper data the mid-infrared excess overprediction 
of z$\sim$2 SFRs when extrapolated from 24~$\mu$m, and motivate the 
discussion of SEDs as a function of offset from the main sequence. 
In Section~\ref{sec:method} we describe the mean-SED redshift-scan fitting 
method. In Section~\ref{sec:by SSFR} we study the mid- to far-infrared 
SEDs as a function of distance from the star forming main sequence, 
demonstrating that this provides a calibration without explicit redshift 
dependence.
We then, in Section~\ref{sec:IRS spectra} compare IRS spectra with the 
modified templates. This also demonstrates that the change in mid-infrared 
SEDs is indeed due to scaling of the relative PAH strength and not due to 
continuum emission from hot dust. In Section~\ref{sec:by LIR} we 
alternatively present the SEDs
in the more traditional way as a function of luminosity and derive a 
redshift-dependent calibration of \nuLnu(8~$\mu$m) as a function of LIR. 
This calibration allows the use of rest-frame 8~$\mu$m photometry as a SFR 
indicator without added knowledge of the mass.
In Section~\ref{sec:AGNs} we discuss mid- to far-infrared SEDs of X-ray AGN 
hosts in the light of the findings for inactive galaxies. 

We adopt a ($\Omega_m$,$\Omega_\Lambda$,$H_0) = (0.3,0.7,70$~km~s$^{-1}$~Mpc$^{-1}$) cosmology throughout this paper.
A \citet{Chabrier03} initial mass function (IMF) is always assumed.
Unless specified otherwise, we refer to the `mid-infrared' as the
rest frame wavelengths corresponding to the MIPS 24~$\mu$m and IRS 16$\mu$m 
blue peakup filters and to the `far-infrared' as the wavelengths covered by the 
PACS 70--160~$\mu$m filters. For star forming galaxies at redshift 
0.7$<$z$<$2.5, these respectively correspond to a region dominated by PAH 
emission features, and by the rest frame far-infrared peak and its short 
wavelength slope.   

%%%%%%%%%%%%%%%%%%%%%%%%%%%%%%%%%%%%%%
\section{Data and samples} \label{sec:Data}
%%%%%%%%%%%%%%%%%%%%%%%%%%%%%%%%%%%%%%

\begin{table}
\begin{center}
\caption{\label{tab:PEP fields} PACS photometry depths in the fields used in this study.}
\begin{tabular}{@{}lccc}
\hline
      & \multicolumn{3}{c}{3$\sigma$ flux limits [mJy]}\\
Field & 70 $\mu$m & 100 $\mu$m & 160 $\mu$m\\
\hline
GOODS-N & \dots & 3.0 & 5.2 \\
GOODS-S & 1.2   & 1.1 & 2.0 \\
\hline
\end{tabular}
\end{center}
\end{table}

\begin{table}
\begin{center}
\caption{\label{tab:Sample_counts} Number and classification of sources in 
the z$\sim$1 and z$\sim$2 samples. The two numbers in each cell refer to 
the GOODS-S/GOODS-N fields.}
%\begin{indented}
\begin{tabular}{@{}lccc}
\hline
          &     & X-ray & Power- \\
Redshift  & SFG & detected$^{\dagger}$ & law$^{\dagger}$ \\
\hline
0.7$<$z$<$1.3 & 148/79 & 21/26 & 1/2 \\
1.5$<$z$<$2.5 & 106/22 & 16/10 & 13/5 \\
\hline
\end{tabular}
\end{center}
$^{\dagger}$ The X-ray detection and IRAC bands power-law tags are not mutually exclusive.\\
\end{table}

\begin{table}
 \begin{center}
  \caption{\label{tab:IRS_statistics} Numbers and luminosity ranges of the 
  full SFG samples and the SFG samples observed with IRS. AGNs are excluded.}
  \begin{tabular}{c|c|c|ccc}
   \hline
   \multicolumn{3}{c}{ } & \multicolumn{3}{c}{ log(LIR$_{8-1000\, \mu{\rm m}}$/L$_\odot$) } \\
   redshifts & Sample & N & Min & Median & Max \\
   \hline
   0.7$<$z$<$1.3 & SFG & 227 & 10.50 & 11.36& 12.2 \\
                 & SFG-IRS & 9   & 11.15 & 11.59& 11.92 \\
   \hline
   1.5$<$z$<$2.5 & SFG & 128 & 11.34 & 12.0 & 12.87 \\
                 & SFG-IRS & 16  & 11.63 & 11.97& 12.72 \\
   \hline
  \end{tabular}

 \end{center}
\end{table}

Our far-infrared data are based on {\it Herschel}-PACS observations in the 
GOODS fields. These were obtained as part of the guaranteed-time PACS 
Evolutionary Probe (PEP\footnote{\url{http://www.mpe.mpg.de/ir/Research/PEP/}}) project. 
For details on the observations field layout, and on data 
reduction we refer to \citet{Lutz11}.
The limiting fluxes are listed in Table~\ref{tab:PEP fields}, for a 
distribution of source LIR versus redshift see Figure 12 of \citet{Lutz11}.
The PACS 160, 100 and 70 $\mu$m fluxes were extracted using sources 
from a {\it Spitzer} MIPS 24~$\mu$m catalog as priors, following the method 
described in \citet{Magnelli09}. The 24~$\mu$m images are very deep 
($\approx$30 $\mu$Jy, 5~$\sigma$) and do not limit the completeness of the 
prior extraction, because only a very small 
fraction ($\lesssim$1\%) of PACS detections will be without a 24~$\mu$m 
counterpart at these depths \citep[see ]
[for a more in-depth analysis]{Lutz11, Magdis11}.
The prior extraction method naturally matches a 
24~$\mu$m source to a PACS source.

For our sample selection we require a redshift (either photometric or 
spectroscopic) and a 160~$\mu$m detection, implying also 24~$\mu$m detection.
The total number of sources and their breakdown by redshift bins and 
SFG/AGN categories are summarized in Table~\ref{tab:Sample_counts}.
To complete the coverage of the mid-infrared at z$\sim$1 we also make use of {\it Spitzer}-IRS blue peak-up 16~$\mu$m catalogs by \citet{Teplitz11}.
The limiting depths (3$\sigma$) in GOODS-N and GOODS-S are 40 and 65 $\mu$Jy, which result in a high detection fraction of PACS sources. In the area covered by the 16 $\mu$m photometry, at redshifts $0.7<z<1.3$, 92\% and 77\% of the PACS sources in GOODS-N and GOODS-S fields respectively have 16~$\mu$m counterparts. 12\% and 6\% of the PACS sources in these respective fields are outside the 16 $\mu$m coverage.

Photometric redshifts in GOODS-N were derived using the code EAZY
\citep{Brammer08} applied on a PSF-matched multi-wavelength catalog in
UV \citep[GALEX,][]{Barger08, Martin05}, U band \citep[KPNO
4m/MOSAIC,][]{Barger08, Capak04}, optical \citep[ACS, HST
bviz,][]{Giavalisco04}, near-IR (FLAMINGOS, JHK$_s$\footnote{Kindly reduced by
Kyoungsoo Lee}) and
K$_s$-band from SUBARU/WIRCam \citep{Wang10}, and mid-infrared ({\it
Spitzer} IRAC and MIPS, Dickinson et al. in prep.).
Spectroscopic redshifts in GOODS-N were adopted from \citet{Barger08}.
In comparison to the z$_{\rm spec}$, the z$_{\rm phot}$ in GOODS-N have a good accuracy with the following statistics for 
$\Delta_z$=(z$_{phot}$ - z$_{spec}$)/(1 + z$_{spec}$): median of -0.008 and median absolute deviation (MAD) of 0.033.
The rate of z$_{\rm phot}$ critical failures, defined here as: $\Delta_z > 0.5\Delta\lambda_{MIPS24}/24$, i.e. a redshift deviation that produces a corresponding wavelength deviation greater than the half-width of the 24~$\mu$m filter is 7\%.
Spectroscopic redshifts in GOODS-S were adopted from \citet{Balestra10} matched to the 
multi-wavelength (bvizJHK, IRAC, MIPS 24~$\mu$m) MUSIC catalog \citep{Santini09} which also includes 
photometric redshifts for the remaining sources.
The corresponding z$_{\rm phot}$ quality statistics for $\Delta_z$ are: median -0.0025, MAD 0.038, critical failures 10\%.
In total, 88\% of the z$\sim$1 and 35\% of the z$\sim$2 samples have spectroscopic redshifts.

Masses were calculated from the above mentioned multi-wavelength catalogs 
of both fields using the method described in 
\citet{Fontana04}, with adjustments as described in \citet{Santini09},
see also \citet{Fontana06} and \citet{Grazian06}.
To ensure good mass estimates when discussing trends with SSFR, we require 
for those samples a detection in all of the 3.6--5.8 $\mu$m IRAC bands 
that cover the SED rest frame stellar bump (rest frame H-band).
Only six 160~$\mu$m sources in our redshift bins do not match this 
criterion, these are included in Tables~\ref{tab:Sample_counts} and
\ref{tab:IRS_statistics} but not used below when masses are required.
We estimate the error on the derived masses by varying the input photometric fluxes according to their respective errors and find that the median error for the galaxies in our sample is 0.2~dex. While this uncertainty is non-negligible on individual objects, it plays a much reduced role in a statistics driven study such as the one presented here. In particular, the masses are used in measuring the galaxy SSFR and its difference from the main sequence, that in itself has a width of 0.3--0.4 dex in SSFR \citep[e.g.,][]{Elbaz07}.

Sources which have an X-ray counterpart in the {\it Chandra} 2~Ms catalogs 
\citep{Alexander03, Luo08} have conservatively been flagged as AGNs.
At a limit of $\sim 2 \times 10^{-17}$ ergs~cm$^2$~s$^{-1}$ and assuming 
the X-ray--SFR relation of \citet{Ranalli03}, the minimum X-ray luminosity of 
$z\sim 2$ galaxies that can be detected is too bright for star formation 
except for extreme SFRs well above those typical for our sample.
This is more of a consideration for the z$\sim$1 sample where the galaxies with the highest SFRs can potentially produce enough X-rays to be detected.
When applying the \citet{Bauer04} classification based on X-ray luminosity and hardness ratio, 18\% of the z$\sim$1 AGN flagged sources may actually be X-ray detected, star forming galaxies. %Figure~\ref{fg:Samples_LIR_vs_z} plots the LIR vs. redshift for the sources flagged as AGNs
%as well as for the star forming galaxies.
Even if 18\% of the AGNs (3\% of the full PACS sample) in the 0.7$<$z$<$1.3 range are star formation dominated in the X-ray, the removal of these sources from the star forming galaxies sample does not have a significant influence. None of the z~$\sim$2 X-ray sources in our sample is suspected to be star formation dominated according to the \citet{Bauer04} classification, though some have very low X-ray photon counts, which makes such a classification highly uncertain.
\citet{Alexander05} investigated luminous LIR$\sim10^{13}$L$_\odot$ SMGs in 
the GOODS-N field. Among 20 SMGs, 15 are X-ray detected with AGN related
X-rays, 3 are not X-ray detected, and only 2 (10\%) are X-ray detected with X-rays
reported to be dominated by star formation. This suggests that at z$\sim$2 the probability to erroneously flag these most luminous star forming galaxies as an AGN, solely based on an X-ray detection in the {\it Chandra} 2~Ms catalog is $\lesssim$10\%. This probability drops sharply towards lower LIRs typical of our 
sample.
Sources that show a monotonic rise in flux in the {\it Spitzer}-IRAC 3.6--8.0 $\mu$m bands are flagged as power-law AGNs. Unless explicitly stated otherwise, the X-ray and power-law AGNs are excluded from the analysis. 

In this study we use the deep {\it Spitzer}-IRS spectra presented by \citet{Fadda10}.
This spectroscopic sample includes 22 sources at $1.75<z<2.4$ and 10 at $0.76<z<1.05$ that are in the PEP GOODS-S field and have a 160~$\mu$m detection. 5 of the z$\sim$2 sources and 1 of the z$\sim$1 sources have an associated X-ray source in the {\it Chandra} 2~Ms catalog.
This sample selects faint 24~$\mu$m sources (0.14-0.5 mJy) and probes the common rest-frame wavelength of 5--12~$\mu$m.
Table~\ref{tab:IRS_statistics} summarizes the numbers and luminosity ranges covered by the full SFG samples and the IRS spectroscopy sub samples.

%%%%%%%%%%%%%%%%%%%%%%%%%%%%%%%%%%%%%%%%%%%%%%%%%%%%%%%%%%%%%%%%%%%%%%%%%%%
\section{The mid-infrared excess: Overprediction of z$\sim$2 SFRs by mid-infrared 
extrapolation} \label{sec:Excess}
%%%%%%%%%%%%%%%%%%%%%%%%%%%%%%%%%%%%%%%%%%%%%%%%%%%%%%%%%%%%%%%%%%%%%%%%%%%

{\it Spitzer}-based findings of overpredicted SFRs based on 
24 $\mu$m photometry at 1.5$<$z$<$2.5  (the `mid-infrared excess' mentioned 
in the introduction) were corroborated and extended by direct measurements of the total infrared luminosity in the far infrared in \citet{Nordon10} and 
\citet{Elbaz10}.
These measurements used GOODS-N imaging with the PACS and SPIRE instruments 
from the {\it Herschel} science demonstration phase.
While resolving many individual sources at these redshifts, they still relied on stacking techniques for a large fraction of the sources: 5\% of PACS detections for the 24~$\mu$m selected sample in \citet{Nordon10} and could not probe luminosities smaller than $10^{12}$~L$_\odot$.
Our current GOODS-S data are about a factor of 3 deeper than the GOODS-N data used before (Tab.~\ref{tab:PEP fields}).
The deeper data significantly increase the 160 $\mu$m detection fraction of 24 $\mu$m sources (30\%), making stacks less necessary. It also allows us to probe down to log(LIR/L$_\odot$)$\approx$11.3, which we could not effectively reach before.

\begin{figure}[t]
 \includegraphics[width=\columnwidth]{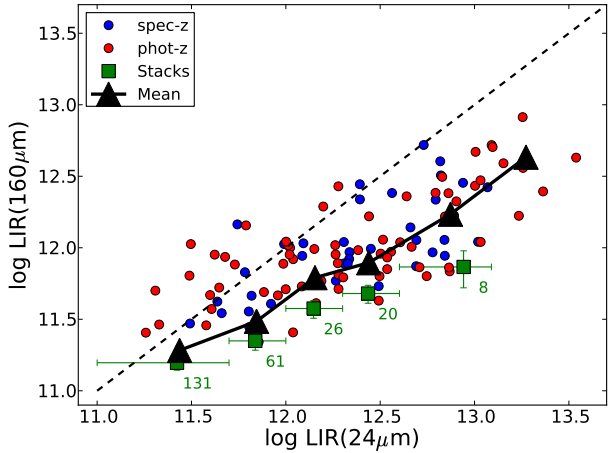}
 % L160_vs_L24_GOODSS.eps: 0x0 pixel, 300dpi, 0.00x0.00 cm, bb=13 175 598 616
 \caption{The `mid-infrared excess' of $1.5<z<2.5$ star forming galaxies in GOODS-S: LIR derived from PACS 160 $\mu$m versus LIR derived from MIPS 24 $\mu$m using the original CE01 templates. Circles indicate individual detections. Squares indicate stacked 24 $\mu$m sources undetected by PACS, where the the number of stacked sources is displayed, the horizontal bars indicate the range of LIR(24) in the stack and the vertical are the error on the mean LIR(160). The number-weighted mean of stacks and detections is plotted in black triangles and thick black line.}
 \label{fg:L160 vs L24}
\end{figure}

In Figure~\ref{fg:L160 vs L24}, using the GOODS-S data, we confirm the results of \citet{Nordon10} and \citet{Elbaz10} (see also upper left panel in Figure~\ref{fg:Compare_Nordon_Murphy_Wuyts_Elbaz} for a different representation).
We plot LIR(160), the LIR as measured from PACS 160 $\mu$m, versus LIR(24) the LIR as extrapolated from MIPS 24 $\mu$m flux using \citet{CE01} SED library. MIPS sources undetected by PACS are stacked in LIR(24) bins. Black triangles are the simple, number-weighted means of stacks and detections in the LIR(24) range of each stack.
All the procedures are similar to those used in \citet{Nordon10}.

The 24 $\mu$m `excess' is clearly evident in Fig~\ref{fg:L160 vs L24}.
At the high luminosities the LIR(24) overestimation is by a factor of $\sim$4, going down to $\sim$2.5 towards log(LIR/L$_\odot$)$\approx$11 which we could not probe directly before.
In the following sections we will demonstrate that this apparent excess is entirely a result of using the wrong SED templates in the 24 $\mu$m flux to LIR conversion.

%% FIGURE1 %%

%%%%%%%%%%%%%%%%%%%%%%%%%%%%%%%%%%%%%%%%%%%%%%%%%%%%%%%%%%%%
\subsection{What determines the mid-to-far IR SED?} \label{sec:iso nuLnu8/LIR}
%%%%%%%%%%%%%%%%%%%%%%%%%%%%%%%%%%%%%%%%%%%%%%%%%%%%%%%%%%%%

Traditionally, the use of mid-infrared photometry as a SFR indicator relied on 
empirical correlations between the mid-infrared spectral features and the 
total infrared luminosity. Since the mid-infrared itself accounts for only 
a small fraction of the 8--1000 $\mu$m LIR, such correlations are not trivial.
If SED shape were strictly related to LIR and given the main sequence 
correlation between stellar mass and LIR, one would expect mass and mid- to
far-infrared SED properties to correlate.
Dependence of SED on other parameters is expected however, as mentioned in 
the introduction.  
As we will demonstrate in this section, there is a dependence of the 
\nuLnu(8)/LIR ratio on both mass and LIR, such that galaxies of constant 
\nuLnu(8)/LIR lie parallel to the main sequence on the SSFR versus mass 
diagram.

Out of our full sample, we consider {\it in this section} only the galaxies with a well 
determined stellar mass and a redshift range of 1.8$<$z$<$2.3 (narrower than the full sample).
At this redshift range the 7.7~$\mu$m PAH complex is inside the MIPS24 filter and the 
10~$\mu$m silicate absorption only affects the filter fluxes to a limited 
degree. Here and consistently throughout the paper, when quoting 
\nuLnu(8)/LIR we refer to the value that would be measured with a MIPS24 
filter for a z=2.0 galaxy of the given SED shape.
We correct the observed MIPS 24~$\mu$m \nuLnu(24/1+z) to 8~$\mu$m rest-frame 
fluxes by finding the CE01 template which best fits the \nuLnu(24/1+z)/LIR of the 
filter and the galaxy redshift, then redshifting this template to z=2 and extracting \nuLnu(8)/LIR 
for the MIPS24 filter. For the subsample used in this section the extrapolation is over a short rest wavelength
interval and the associated uncertainties are negligible.

\begin{figure}[t]
 \includegraphics[width=\columnwidth]{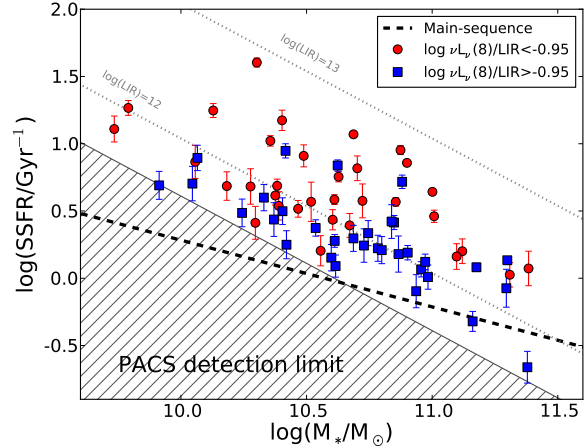}
 % nuLnu8LIR_on_SSFR_diagram_z2.eps: 0x0 pixel, 300dpi, 0.00x0.00 cm, bb=13 175 598 616
 \caption{SSFR versus mass for 1.8$<$z$<$2.3 galaxies. 
Galaxies with log(\nuLnu(8)/LIR)$<$-0.95 are in red and galaxies with 
log(\nuLnu(8)/LIR)$>$-0.95 are in blue. The dashed line is the main sequence 
of \citet{Rodighiero10} with a slope of -0.5.
Gray dotted lines with a slope of -1 indicate lines of constant LIR.
The gray hatched area is 
excluded by the PACS 160 $\mu$m detection limit for the GOODS-S sample.}
 \label{fg:nuLnu8LIR SSFR diagram z2}
\end{figure}

Figure~\ref{fg:nuLnu8LIR SSFR diagram z2} plots this subsample on the SSFR 
vs. M$_*$ diagram. Red and blue colors indicate galaxies with low $<$-0.95 
and high $>$-0.95 \nuLnu(8)/LIR ratios, which splits the sample roughly
in half. While the separation is not clean, it 
is clear that sources with low \nuLnu(8)/LIR tend to have higher SSFR at any 
given mass.
Probing by eye the somewhat blurred border between the blue and red 
points, it appears to be sloped rather than at a horizontal line
of constant SSFR.
Lines of constant LIR appear in this plot at a slope of -1 (gray dotted lines).
The blue/red separation is not as steep as these lines of constant LIR.
The main sequence from \citet{Rodighiero10}, which 
is based on {\it Herschel}-PACS GOODS-N data is plotted as a dashed line and its intermediate slope
of $\sim$-0.5 appears to be approximately parallel to the blue/red 
(low/high \nuLnu(8)/LIR) separation line.
In this interpretation, galaxies with constant \nuLnu(8)/LIR thus appear along this mass-dependent 
SSFR(M$_*$) of the main sequence, rather than lines of constant SSFR or constant LIR.
While lines of \nuLnu(8)/LIR seem to be closest to the 
slope of the main sequence, in practice over this limited mass range and the scatter in \nuLnu(8)/LIR,
binning the galaxies with respect to the main sequence (slope -0.5), constant 
SSFR lines (0 slope), or constant LIR lines (-1 slope) will not produce 
drastically different selections in terms of \nuLnu(8)/LIR.

\begin{figure*}[t]
\begin{center}
 \includegraphics[width=0.85\textwidth]{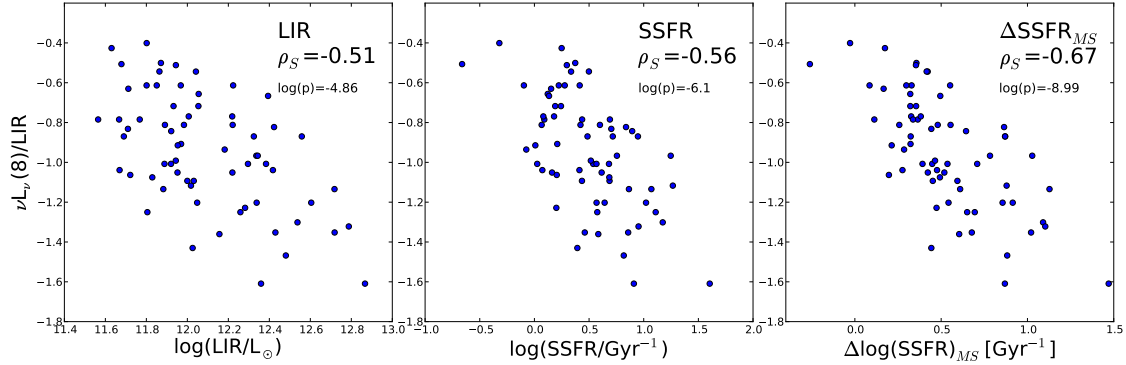}
 % Spearman_corr_tests_z2.eps: 0x0 pixel, 300dpi, 0.00x0.00 cm, bb=-338 189 950 602
 \caption{Spearman rank correlation test of \nuLnu(8)/LIR against LIR, SSFR and SSFR-distance from the main sequence. The correlation coefficient $\rho_s$ and the null-hypothesis probability (p-value) are indicated in each panel.}
 \label{fg:Spearman corr test z2}
\end{center}
\end{figure*}

We quantify the latter statement via a Spearman rank correlation test 
of \nuLnu(8)/LIR against LIR, SSFR and the offset from the main sequence
at the given galaxy mass $\Delta\log({\rm SSFR})_{\rm MS}=\log({\rm SSFR})-\log({\rm SSFR})_{\rm MS}$.
The results are plotted in Figure~\ref{fg:Spearman corr test z2}. All three 
tests reject the null hypothesis of uncorrelated data with high 
significance, but the best correlation coefficient and highest significance
of the three is with the distance from the main sequence. We note that in
case of a perfect correlation of \nuLnu(8)/LIR and 
$\Delta$log(SSFR)$_{\rm MS}$, significant correlation would also be expected
with LIR and SSFR over the range studied, given the definitions of these
quantities.

We thus found a tendency of galaxies with similar SEDs to lie along sloped 
lines in the SSFR versus mass diagram. This tendency corresponds to the
best correlation found between \nuLnu(8)/LIR and $\Delta$log(SSFR)$_{\rm MS}$.
 It is consistent with the 
\citet{Rodighiero10} main sequence slope, within the statistics
of this small sample. The slope of these lines as opposed to a 
constant SSFR makes also the correlation of \nuLnu(8)/LIR with LIR 
significant, even though it is 
not the best relation of the three. The physical origin of the correlation between SED shape and LIR 
hence seems more indirect, not fundamental as we shall discuss further in 
Section~\ref{sec:Discussion}. Still, for practical purposes this correlation 
can be used to measure LIR from 24~$\mu$m fluxes in cases where the mass is 
not well determined, or unknown.

In the following sections we will explore the mid-to-far IR SEDs in two 
ways: in the distance from the main sequence that seems to be 
the more fundamental relation, and the classical and simpler dependence on LIR.

%%%%%%%%%%%%%%%%%%%%%%%%%%%%%%%%%%%%%%%%%%%%%%%%%%%%%%%%%%%%%%%%%%%%%%%%
\section{Method: mean SED redshift scan} \label{sec:method}
%%%%%%%%%%%%%%%%%%%%%%%%%%%%%%%%%%%%%%%%%%%%%%%%%%%%%%%%%%%%%%%%%%%%%%%%

The study of galaxy SEDs in the mid-to-far IR range can be broken into two 
parts: the {\it shape} of the SED and the {\it overall amplitude}, i.e. the 
total infrared luminosity (LIR), defined as the total integrated luminosity 
between 8--1000~$\mu$m.

For studying the SED shape, one must overcome a few difficulties. The first is that due to various reasons, many of the galaxies are not detected at all wavelengths. The non-detections must be taken into account in order not to bias the results. The second is that at different redshifts the filters sample different rest-frame wavelengths. Normally this requires some inter/extrapolation to common wavelengths, that require assumptions on the same region of the SED we wish to study. This is most problematic in the mid-infrared where there are strong spectral variations on scales smaller than the filter width.

We wish to study the {\it mean} properties of mid-to-far IR emission. Therefore, instead of fitting a template SED to every individual galaxy, we take a slightly different approach:  we fit a template to the combined photometric data points of a galaxy population in their rest-frame.
This treats the combined data as if originating from a single {\it mean} galaxy, observed by many different filters which are slightly shifted in their central wavelengths.
In this method, sampling a narrow, yet non-negligible redshift interval requires no interpolation/extrapolation to a common rest frame-wavelength.
Such extrapolations can be problematic in the mid-infrared due to PAH features.
It also does not force the SEDs to cross at a single wavelength chosen for the normalization and the sampling of slightly different rest-frame wavelengths by the filters improves the effective resolution.

The first and most important piece of information we need for each galaxy is its LIR. 
{\it Herschel}-PACS, especially with the 160~$\mu$m filter, allows us to accurately determine the LIR almost independently of the SED shape.
Generally, in our fields and at the redshifts of interest, out of the 3 PACS bands, 160~$\mu$m has the highest detection rate.
It also samples close to 60 $\mu$m rest-frame at z$\sim$2, which leads to a very robust LIR determination as \citet{Elbaz10} demonstrated. 
60 $\mu$m rest-frame is a `sweet spot' for monochromatic LIR determination and we expand on this point further in Appendix~\ref{app:determin LIR}.
For the 1.5$<$z$<$2.5 sample we measure LIR by fitting CE01 templates to 160 $\mu$m flux alone, while preserving the library LIR-template relation (monochromatic fitting), even in cases where shorter wavelength fluxes are available.
At 0.7$<$z$<$1.3 the PACS 160 $\mu$m filter has moved farther away from the rest-frame 60 $\mu$m sweet-spot and PACS 100 $\mu$m filter has moved closer.
For galaxies at these lower redshifts we fit the best combination of CE01 template shape and scale (two free parameters) to the 100 and 160 $\mu$m points simultaneously, except for a few cases at the lowest luminosities where only 160 $\mu$m is available.
In the latter case a monochromatic fit to LIR is used.
For further discussion on the LIR determination we refer the reader to Appendix~\ref{app:determin LIR}.

After determining the LIR of each galaxy in our full sample, we are able to 
bin the sample by SSFR offset from the main sequence or by LIR. For each bin 
subsample we fit a single template to the \nuLnu/LIR values of all sources 
combined. Dividing all filter luminosities of a given galaxy by its 
own LIR serves as normalization and does not affect the flux ratios. 
Each galaxy thus contributes nearly equally to the overall \nuLnu/LIR template fit,
only depending on its photometry errors, but independent of its absolute luminosity.
As a working example, we start by selecting $1.5<z<2.5$ PACS 160~$\mu$m sources according to their $\Delta$log(SSFR)$_{\rm MS}$ distance from the main sequence in the SSFR-M$_*$ plane.
LIR is converted to SFR using the relation from \citet{Kennicutt98}, converted to a Chabrier IMF \citep{Chabrier03}:
\begin{equation}
 \left( \frac{SFR}{M_\odot yr^{-1}} \right) = 1.09\times 10^{-10} \left( \frac{LIR}{L_\odot} \right)
 \label{eq:LIR2SFR}
\end{equation}

The exact parametrization of the main sequence of star forming galaxies is 
somewhat ambiguous. Different studies (at different wavelengths and using 
different selections of star forming galaxies) find different slopes $n$ 
and a corresponding scaling $\Omega_{MS}$ for the relation:
\begin{equation}
 \log \left( \frac{\dot{M}_*}{M_*} \right) = n \log \left( \frac{M_*}{M_\odot} \right) + \log(\Omega_{MS})
\end{equation}
In this study we use the main sequence as determined by \citet[][corrected to a Chabrier IMF]{Rodighiero10} that was derived from PEP data in GOODS-N:
$(n, \log(\Omega_{MS})) = (-0.496,5.243)$ for z$\sim$2 and $(n, \log(\Omega_{MS})) = (-0.394,3.945)$ for z$\sim$1.
The redshift of 1 is between two bins in \citet{Rodighiero10} and we adopt the simple mean of the main-sequence above and below it.

\begin{figure}[t]
 \includegraphics[width=\columnwidth]{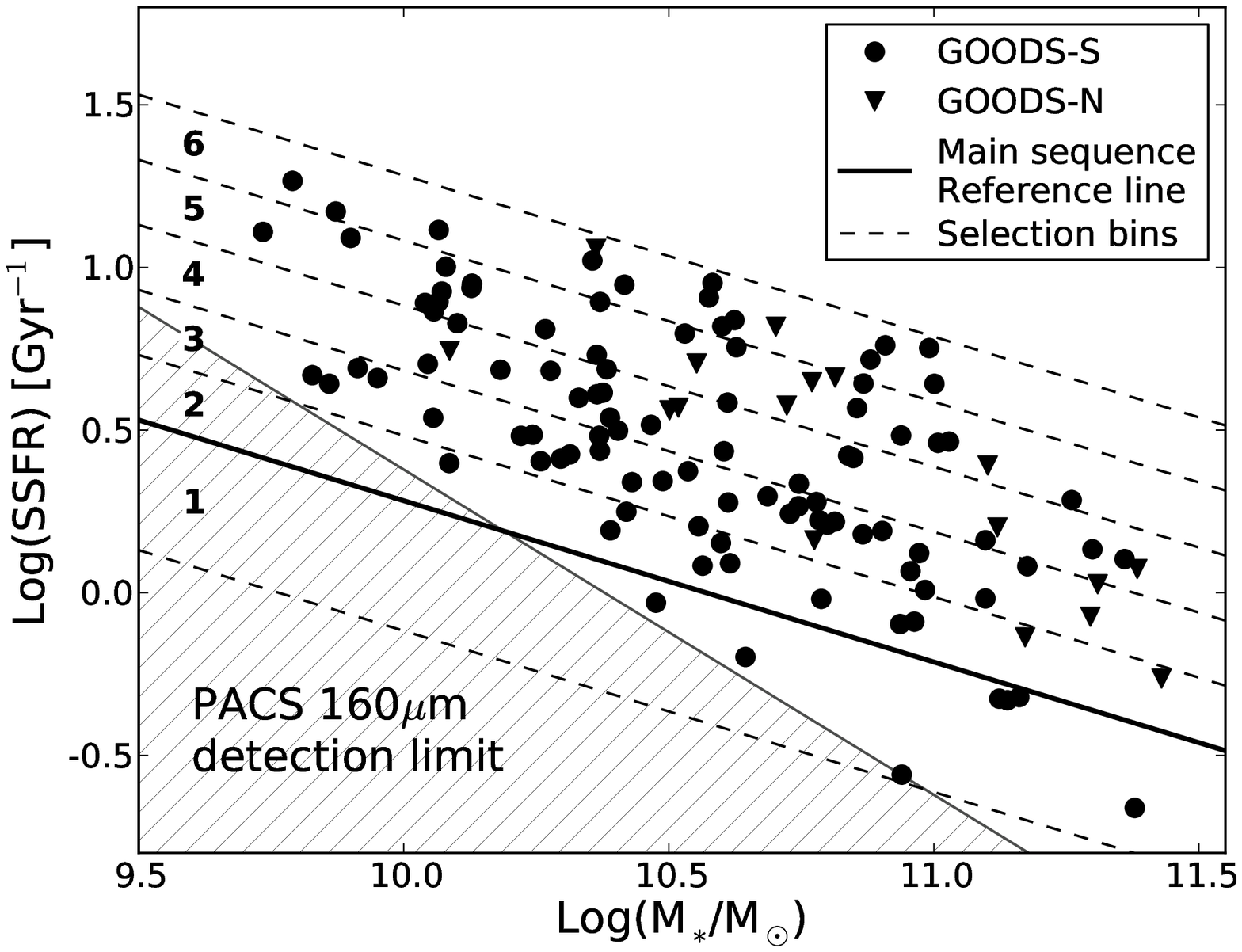} \\
 \includegraphics[width=\columnwidth]{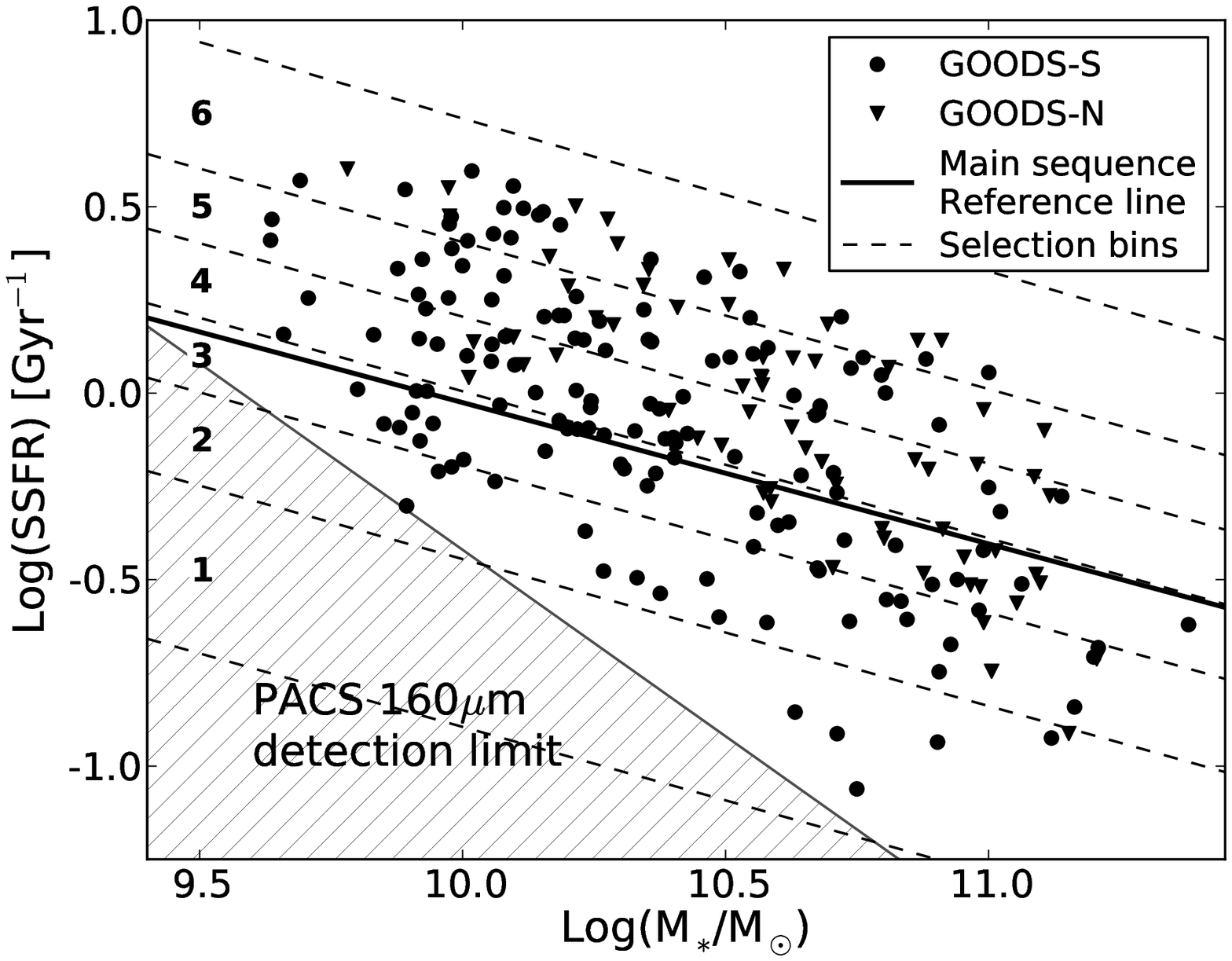}
 \caption{{\it Top:} The binning of the $1.5<z<2.5$ sample according to the distance from the main sequence.
 {\it Bottom:} Similar to above for the $0.7<z<1.3$ sample.}
 \label{fg:bins by dSSFR}
\end{figure}

\begin{figure*}[t]
\begin{center}
 \includegraphics[width=0.99\textwidth]{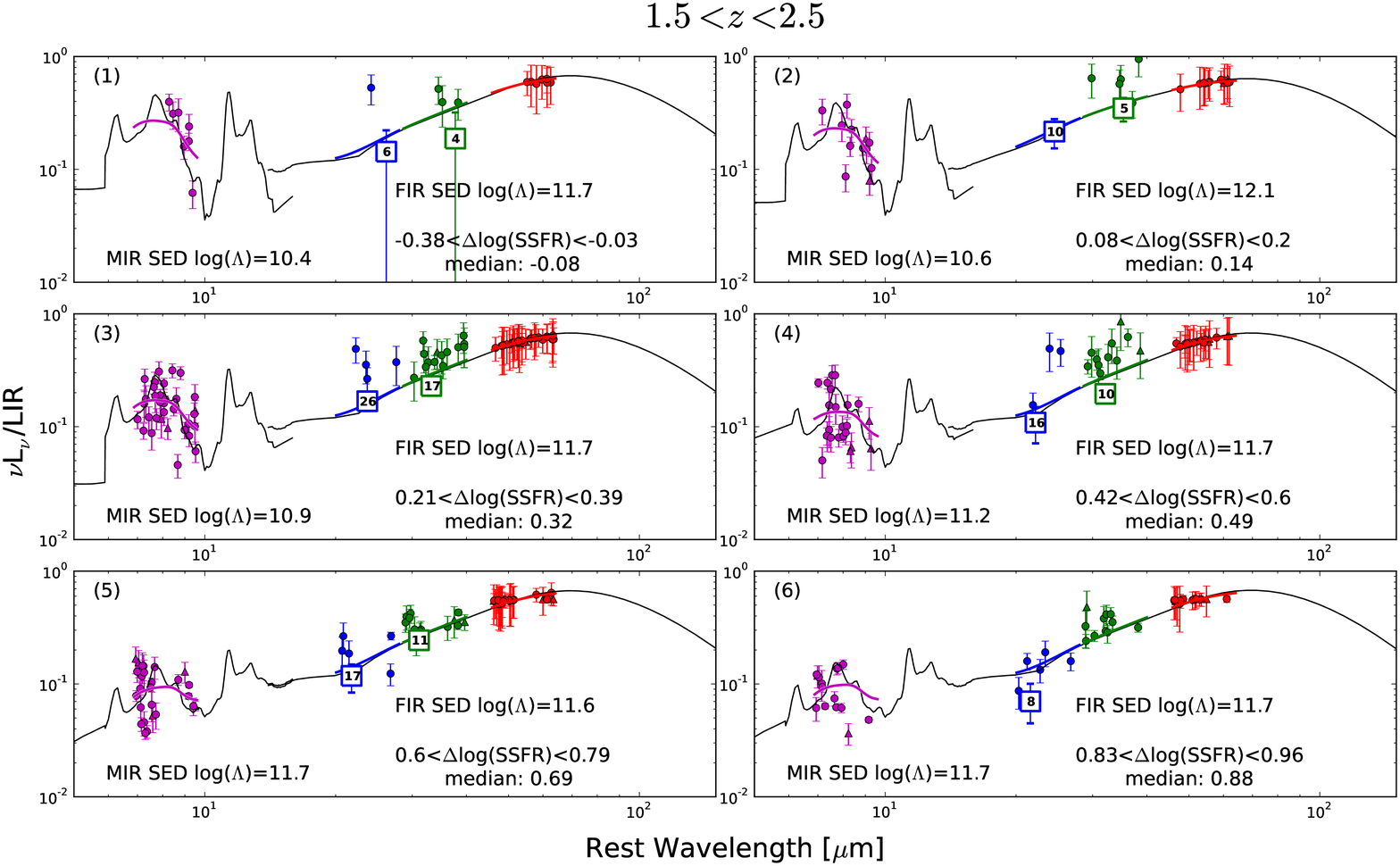} \\
 \includegraphics[width=0.99\textwidth]{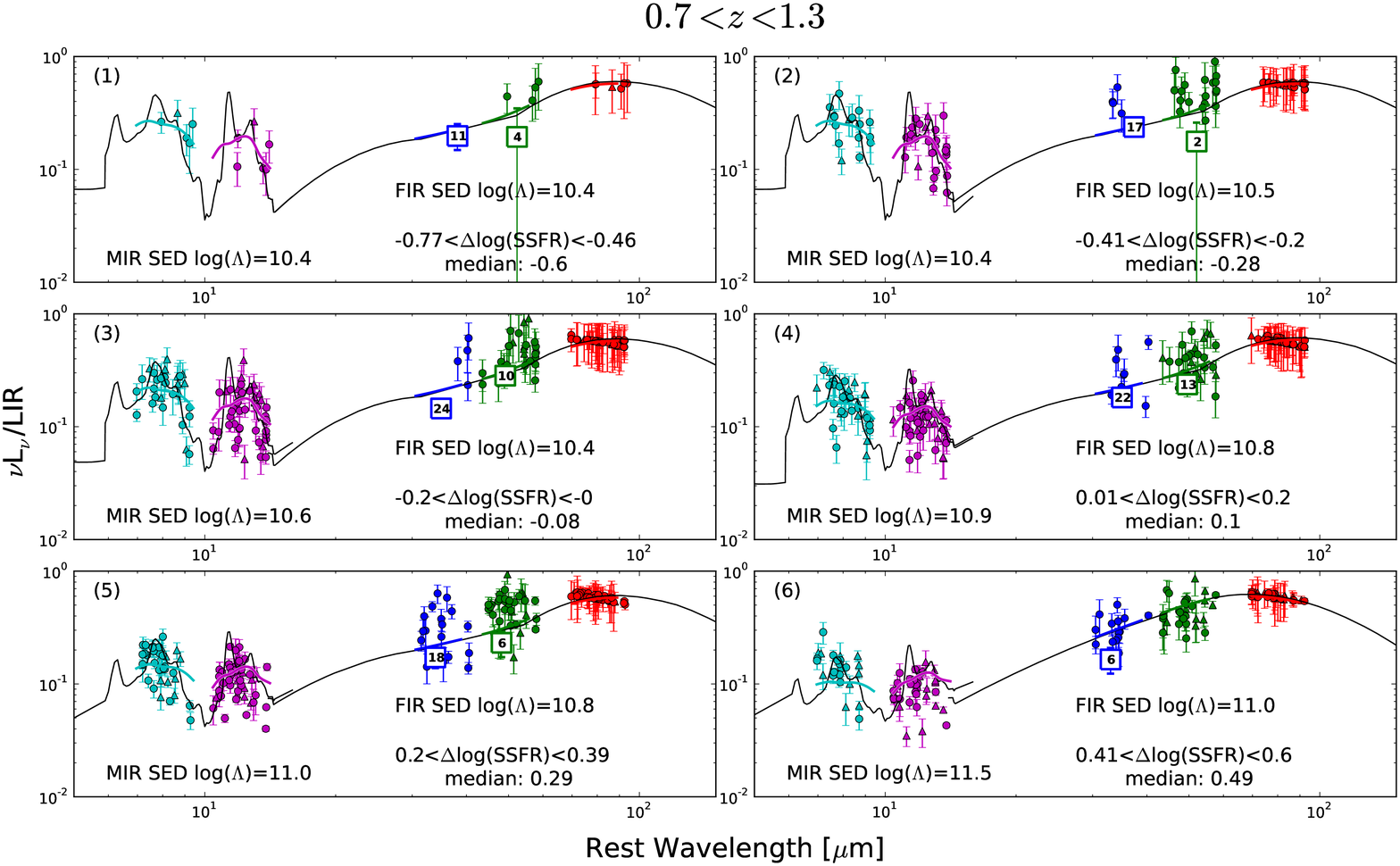}
 \caption{ {\it Top:} Mean SEDs fitted to the $1.5<z<2.5$ sample, binned by the $\Delta$log(SSFR)$_{\rm MS}$  distance from the main sequence line. The number on the corner of each panel indicate the corresponding  $\Delta$log(SSFR)$_{\rm MS}$ bin in Figure~\ref{fg:bins by dSSFR} top panel. Circles and triangles are for 
 GOODS-S and GOODS-N sources respectively. Squares mark stacks of GOODS-S 
 undetected sources with the number of stacked sources indicated on them. The colors: red, green, 
 blue and purple, correspond to the filter wavelengths: PACS 160, 100, 70 
 and MIPS 24~$\mu$m. CE01 library SED template shapes \nuLnu/LIR are fitted 
 to the sample separately for the mid- and far-infrared rest wavelength ranges below and 
 above 15~$\mu$m. These fits are plotted in solid black and the best fitting 
 SED shapes are labeled by their nominal luminosity $\Lambda$.
 The $\log(\Lambda)$ template identifier of the best 
 fitting templates are noted separately for the mid- and far-infrared.
 The expected filter fluxes for the best-fit template are plotted as thick lines matching 
 the filter color.
 {\it Bottom:} Same as above for the $0.7<z<1.3$ sample. The additional 16~$\mu$m photometry is colored in cyan. The panel numbers match the bins plotted in Figure~\ref{fg:bins by dSSFR} bottom panel.
  }
 %\label{fg:CE01nuLnuLir z2 SSFR}
 \label{fg:nuLnuLir fit by SSFR}
\end{center}
\end{figure*}

We divide our $1.5<z<2.5$ sample into 6 bins as illustrated in 
Figure \ref{fg:bins by dSSFR} top panel. 
The bins are numbered 1--6 with increasing $\Delta\log({\rm SSFR})_{\rm MS}$.
Also indicated is the region below a line with slope -1 that is excluded 
by the 160~$\mu$m flux limit.
The combined photometry for each $\Delta$log(SSFR)$_{\rm MS}$ bin is plotted in Figure~\ref{fg:nuLnuLir fit by SSFR} top.
Each filter flux is converted to \nuLnu/LIR units according to the corresponding galaxy LIR and the data points are color coded according to their filter.
Stacks of (individually) undetected 100 and 70~$\mu$m (but detected at 160~$\mu$m) sources are also plotted in the respective filter color.
On top of each stack point we indicate the number of sources in the stack and the error bars indicate the uncertainty on the {\it mean} \nuLnu/LIR in the stack.
For details about the stacking procedure see Appendix~\ref{app:stacking}.

An SED template is then fitted to the combined photometry of all galaxies in each bin using $\chi^2$ minimization.
Since we are not making repeated measurements of the same galaxy, but rather of a galaxy population, the intrinsic scatter in the population must be taken into account when calculating $\chi^{2}$.
This has a significant effect on the relative weight of each photometric point and the relative weight of the stacks.
More details about the $\chi^{2}$ minimization are available in Appendix~\ref{app:chisqr_minimization}.
We would like to keep the description of the SED shapes as simple as possible.
Due to its popularity, its use in determining LIR and the simple way in which it is defined, we choose to use the CE01 library. \citet{Magnelli09} also found that at z$\sim$1 this library reproduce the 24/70 micron colors better than a few other popular libraries.
Even though we concentrate on CE01, we will supply the means to calibrate any other library.
The templates in the CE01 library represent the mid-infrared to sub millimeter SEDs
of local galaxies according to their total IR luminosity, and are uniquely 
identified by their LIR. 
We also use them below as SED shape templates that are encoding relative 
ratios between the far- and the mid-infrared, and the PAH features in particular, without 
enforcing the locally calibrated link to total IR luminosity. 
To avoid confusion, we refer in this case to the template {\it identifier} 
as $\Lambda_{\rm CE01}$ in units of L$_\odot$, irrespective of the IR luminosity 
to which it might have been rescaled.

We thus search the CE01 library for the best fitting template {\it shape}. 
Since both data and templates are in \nuLnu/LIR units, there is no scaling 
involved in this fit. We split at 15~$\mu$m rest-frame and fit the mid- and 
far-infrared independently.
This split differentiates two distinct wavelength and physical regions: The 
mid-infrared is dominated by PAH emission from cold dust regions, silicate absorption 
and possible continuum from AGN-heated hot dust, while the far-infrared 
peak is dominated by graybody emission of dust. In the middle, in the range 
of 15--60~$\mu$m (probed by PACS), a tail distribution of hotter 
dust temperatures and transient heating of small grains shape the SED slope
\citep[e.g.][]{Desert90}.

In each panel of Figure~\ref{fg:nuLnuLir fit by SSFR} we indicate the best fitting mid- and far-infrared $\Lambda_{CE01}$ templates. The best fitted SED is plotted as a solid black line.
Broadband filter fluxes represent the weighted emission from a significant wavelength interval. 
For each filter we indicate with a solid colored line the expected filter flux for the fitted template by gradually redshifting the fitted template in the range 1.5$<$z$<$2.5 and applying the filter transmission to the template.
The central wavelength of the filter scans a corresponding range of rest-frame wavelengths (redshift-scan).
The distinction between the template and the curve of expected filter flux is most important in the mid-infrared region.

One final note should be made: Since we require 
PACS 160~$\mu$m detections in order to estimate LIR, sources 
undetected in this filter are excluded. We have argued above and in Appendix~\ref{app:determin LIR} that 160~$\mu$m
is a good proxy for LIR at these redshifts over a wide range of SEDs.
When later binning by LIR (Section~\ref{sec:by LIR}), non-detections at 160~$\mu$m thus do not introduce significant biases as we operate in narrow LIR bins: the near
independence of rest frame \nuLnu(60~$\mu$m)/LIR from SED shape means that 
non-detections are mostly due to limiting luminosity-distance for a given LIR.
When binning by $\Delta$log(SSFR)$_{\rm MS}$, this translates into a more complex combination of limiting luminosity-distance and M$_*$. Unless there is an added dependence of SED shape on the absolute mass, which is not a result of the SFR--M$_*$ relation, this binning does not introduce biases as well.

%%%%%%%%%%%%%%%%%%%%%%%%%%%%%%%%%%%%%%%%%%%%%%%%%%%%%%%%%%%
%
\section{The mid-to-far IR SED as a function of distance from the main sequence} \label{sec:by SSFR}
%
%%%%%%%%%%%%%%%%%%%%%%%%%%%%%%%%%%%%%%%%%%%%%%%%%%%%%%%%%%%%
\begin{figure*}[t]
\begin{center}
 \includegraphics[width=0.75\textwidth]{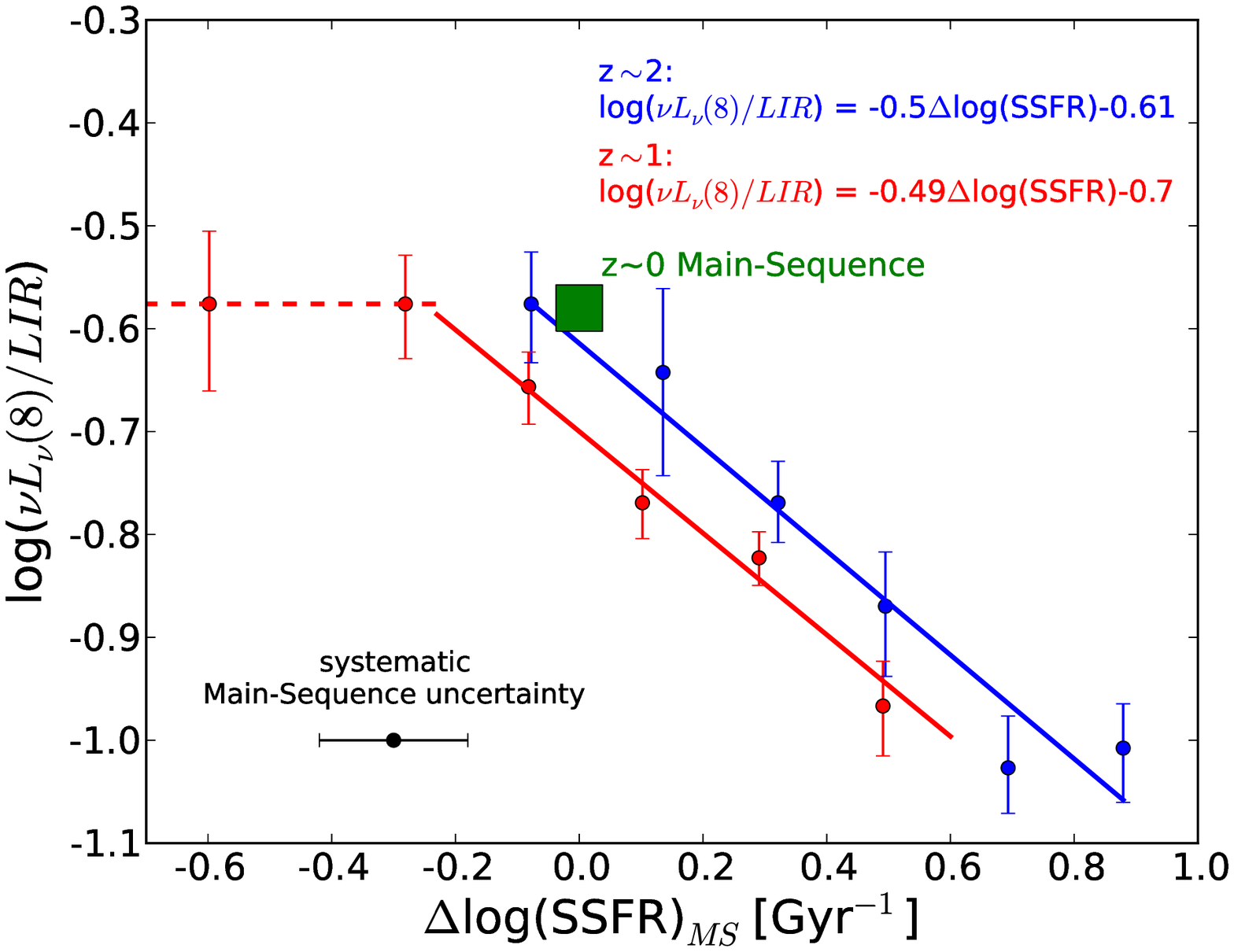}
 \caption{\nuLnu(8 $\mu$m)/LIR vs. the $\Delta \log({\rm SSFR})_{\rm MS}$ 
specific star formation rate offset from the main sequence for the two redshifts samples. The green square indicates z$\sim$0 main sequence galaxies.
 The systematic uncertainties on $\Delta \log({\rm SSFR})_{\rm MS}$ due to the placement of the main sequence are indicated on the lower left. The fitted relations use $\chi^2$ minimization.
 \nuLnu(8) value is the MIPS24 filter luminosity as would be measured for a galaxy at z=2.}
 \label{fg:CE01 rescale SSFR}
\end{center}
\end{figure*}

In Section~\ref{sec:iso nuLnu8/LIR} we found that galaxies with similar 
\nuLnu(8)/LIR tend to lie parallel to the main sequence in their SSFR,
and Section~\ref{sec:method} described our method to build up mean SEDs for
z$\sim$2 galaxies as a function of offset from the main sequence.
We follow a similar procedure for our $0.7<z<1.3$ sample. At z$\sim$1 the 
MIPS24 filter samples rest-frame wavelengths $\lambda \sim 12$~$\mu$m, 
which cover entirely different spectral features than those at 8~$\mu$m.
We therefore add the 16~$\mu$m IRS Blue Peak-up filter data from 
\citet{Teplitz11} that samples 8 $\mu$m rest-frame.
The sample is divided into 6 $\Delta$log(SSFR)$_{\rm MS}$ bins as 
illustrated in Figure \ref{fg:bins by dSSFR} bottom panel and SEDs are again fitted to the combined 
photometry of all galaxies in each bin.
The fit results are plotted in Figure \ref{fg:nuLnuLir fit by SSFR} bottom.

For both z$\sim$2 and z$\sim$1 (Figure \ref{fg:nuLnuLir fit by SSFR} top and bottom respectively),
in the sequence of increasing bin number and SSFR, there is a clear trend towards fitting the
mid-infrared ratio by template shapes corresponding to higher
luminosities $\Lambda_{CE01}$, implying lower $\log({\nu}L_{\nu}(8)/LIR)$.
In the next section (\ref{sec:building SSFR scaled library}) we will quantify these effects.

In contrast, the far-infrared $\lambda>15$~$\mu$m is 
fitted by similar templates in all six bins of the z$\sim$2 sample. 
In the CE01 library all templates in the range 
11$<$log($\Lambda$/L$_\odot$)$<$12.1 are nearly identical in the rest frame 
wavelengths covered by the 3 PACS bands
(20--60~$\mu$m, see also Figure~\ref{fg:nuLnuLirLib} in Appendix~\ref{app:determin LIR})
when plotted in \nuLnu($\lambda$)/LIR scale.
This part of the far-infrared SED, among this range of 
templates, is fairly insensitive to the exact location of the SED peak. 
For the z$\sim$2 sample we hence
cannot reliably infer dust temperatures and which template will be the 
most accurate in extrapolating to longer rest-frame wavelengths.
We note, however, that the templates fitted to the mid-infrared 
($\lambda<15$~$\mu$m) also fit the longer wavelengths up to 60~$\mu$m at 
z$\sim$2. For the z$\sim$1 sample, the best-fit templates for the 
far-infrared up to $\sim$90~$\mu$m rest-frame are close to the templates 
fitted to the mid-infrared, as indicated by their respective $\Lambda_{CE01}$.
Within the degeneracy of the templates we thus find consistent fit results
for the mid- and far-infrared, but do not strongly constrain the shape of 
the rest far-infrared peak given the rest wavelength coverage of the PACS 
data.

%%%%%%%%%%%%%%%%%%%%%%%%%%%%%%%%%%%%%%%%%%%%%%%%%%%%
\subsection{Building a SSFR-scaled SED library} \label{sec:building SSFR scaled library}
%%%%%%%%%%%%%%%%%%%%%%%%%%%%%%%%%%%%%%%%%%%%%%%%%%%%

We wish to derive a recipe for calibrating the CE01 SEDs by offset from 
the main sequence.
Figure~\ref{fg:CE01 rescale SSFR} concentrates all the essential mid-infrared information from the template fittings described above into one figure and presents our main result.
It plots the \nuLnu(8 $\mu$m)/LIR versus the $\Delta$log(SSFR)$_{\rm MS}$ deviation from the main sequence.
For each template fitted to the mid-infrared photometry in Figure~\ref{fg:nuLnuLir fit by SSFR} we derive the mean \nuLnu(8)/LIR as observed through the MIPS 24 $\mu$m filter at z=2 (centered on 8~$\mu$m rest frame).
This scaling curve can be used to recalibrate any SED library (Appendix~\ref{app:calibrate library}).
Filter convolved flux is used instead of the monochromatic
template point at 8~$\mu$m because our template calibration is photometric, 
in a structured region of the SED.
We cannot be sure that the template is correct in its higher resolution details, which also may 
vary in other libraries, but note good agreement of PAH dominated CE01 
template shapes with IRS spectra for a subset of our sample 
(Section~\ref{sec:IRS spectra}).
For z$\sim$1, even though the 8~$\mu$m was observed by the 16~$\mu$m filter, 
we again specify \nuLnu(8)/LIR in terms of the z=2.0 MIPS 24~$\mu$m flux for 
the template.

The resulting curves of \nuLnu(8)/LIR as a function of 
$\Delta$log(SSFR)$_{\rm MS}$ for the two redshifts bins are plotted in 
Figure~\ref{fg:CE01 rescale SSFR}.
The slopes of the z$\sim$2 and z$\sim$1 samples for $\Delta$log(SSFR)$_{\rm MS} \gtrsim 0$ are practically identical and the curves agree within the data uncertainty (plotted error bars), and in particular the added systematic ambiguity of placement of the main sequence itself at different redshifts.
For our adopted main sequence from \citet{Rodighiero10} we estimate
this placement to carry an uncertainty of $\sim$0.1--0.15~dex, also 
considering slightly different methods to derive LIR.
This main sequence ambiguity results in horizontal shifts of the curves and can account for the $\Delta$log(SSFR)$_{\rm MS}$ shift between the z~$\sim$1 and z$\sim$2 curves.

Galaxies which are on the main sequence at the two redshift bins have nearly the same \nuLnu(8 $\mu$m)/LIR ratio, and this ratio changes with the distance from the main sequence in a similar fashion.
More importantly, it means that the increase with redshift of the SSFR at 
a given mass (i.e. the scaling of the main sequence with redshift), is 
the same as the scaling that affects SED shapes.

The clear negative slope indicates a suppression of 8~$\mu$m emission vs. 
the LIR for galaxies with SSFR higher than the main sequence. Quantitative 
details of this result are sensitive to the slope and in particular 
scaling of the adopted main sequence which vary between different studies.
Adopting a different main sequence will change the absolute calibration derived here, but not the fundamental behavior, namely the saturation of log(\nuLnu(8)/LIR)=-0.58 on/below the main sequence. It will hardly change the slope of the curves in Figure~\ref{fg:CE01 rescale SSFR}.

For objects with SSFR below the main sequence, which are available only in the 
z$\sim$1 sample, log(\nuLnu(8)/LIR) saturates into a constant value of -0.58$\pm$0.04.
This is the typical (near constant) value for templates in the CE01 library 
for local galaxies of log(LIR/L$_\odot$)$<$10.2.
In the CE01 library log(\nuLnu(8)/LIR) was constrained by comparing 
ISOCAM-LW2 6.7~$\mu$m (also covering the 7.7 $\mu$m PAH band) with the total infrared luminosity
\citep[Figure 3 in ][]{CE01}.
The correlation is almost linear up to log(LIR/L$_\odot$)$\approx$10 and then breaks towards progressively reduced
\nuLnu(6.7)/LIR  with increasing LIR. A characterization of
log(\nuLnu(8)/LIR) as a function of main sequence offset 
$\Delta$log(SSFR)$_{\rm MS}$ for local galaxies is beyond the scope of this 
work, but we can safely assume that the normal galaxies below the log(LIR/L$_\odot$)$\approx$10 break point for \nuLnu(8)/LIR in CE01 are main 
sequence star formers. We use their log(\nuLnu(8)/LIR) to 
indicate the location of the z=0 main sequence in 
Figure~\ref{fg:CE01 rescale SSFR} (green square).
The break-point luminosity translates into a SFR of 
$\sim$1~M$_\odot$~yr$^{-1}$, which at z=0 means $\sim$10$^{10}$~M$_\odot$ 
main sequence galaxies \citep{Brinchmann04,Peng10}. Due to the specifics of 
the CE01 sample, higher luminosities are mostly sampled by very luminous 
z=0 ULIRG mergers for which 8~$\mu$m emission is suppressed. With masses of a 
few 10$^{10}$M$_\odot$ \citep[e.g.,][]{Dasyra06} these ULIRGs will be 
clearly above the local main sequence. This implies a downward trend in 
Figure~\ref{fg:CE01 rescale SSFR} at positive SSFR offsets above the main 
sequence, also for z=0.
\citet[][their Equation~5]{Elbaz11} find consistent values of 
LIR/\nuLnu(8)=4 in local main sequence galaxies, and that this ratio holds 
for main sequence galaxies up to z$\sim$2.5.

For the z$\sim$2 sample in Figure~\ref{fg:CE01 rescale SSFR} we 
could not resolve the break point between sloped log(\nuLnu(8)/LIR) and 
the likely constant value at lower  $\Delta$log(SSFR)$_{\rm MS}$. This is 
due to our inability to include galaxies further below the main sequence at 
this redshift.
The lowest $\Delta$log(SSFR)$_{\rm MS}$ point which is on and slightly below 
the main sequence is consistent with the log(\nuLnu(8)/LIR)=-0.58 limit 
derived from z$\sim$1 and may already include galaxies past the break.

Figure~\ref{fg:CE01 rescale SSFR} implies that there is a single relation between
SED shape (i.e. \nuLnu(8)/LIR and the matching 8--60 $\mu$m template in general) and the SSFR offset from the main sequence that applies all the way 
from z$\sim$0 to z$\sim$2.
This relation can be expressed in terms of \nuLnu(8)/LIR:

\begin{equation}
 \begin{array}{ll}
  \log({\nu}L_{\nu}(8)/LIR) = \\
  -0.58_{\pm 0.04}                                & : \Delta_{\rm MS} < -0.14\\
  -0.65_{\pm 0.03} - 0.50_{\pm 0.06}\times\Delta\log({\rm SSFR})_{\rm MS}& : \Delta_{\rm MS} \geq -0.14
 \end{array}
 \label{eq:nuLnu8/LIR vs SSFRoffset}
\end{equation} 

\noindent where $\Delta_{\rm MS} \equiv \Delta\log({\rm SSFR})_{\rm MS}$ is expressed in Gyr$^{-1}$ units and we adopt the mean from z$\sim$1 and z$\sim$2 for the sloped part.
The errors on the fitted slope and intercept coefficients are correlated with $\rho=-0.75$.
For the specific case of the CE01 library, it can be approximated by 
luminosity-labels $\Lambda_{CE01}$ of this library:

\begin{equation}
 \begin{array}{ll}
  \log(\Lambda_{\rm CE01}) = \\
  10.38                                            & : \Delta_{\rm MS} < -0.14\\
  10.58 + 1.44\times\Delta\log({\rm SSFR})_{\rm MS}& : \Delta_{\rm MS} \geq -0.14
 \end{array}
 \label{eq:LCE01 vs SSFRoffset}
\end{equation} 

We find a consistent decrease of the mid-to-far infrared ratio 
\nuLnu(8)/LIR in the SED for galaxies that are offset above the main sequence.
The strong scaling of SFR on the main sequence with redshift 
SFR/M$_*\propto (1+z)^{2.7}$ \citep{Bouche10} makes LIRG and ULIRG-like 
luminosities common 
for massive star forming galaxies on the main sequence at z$\sim$2.
Most of these highly star forming galaxies indeed have IR SED shapes that 
resemble 
lower redshift main sequence galaxies. For both z$\sim$1 and z$\sim$2, a 
decrease in \nuLnu(8)/LIR is observed with increasing SSFR above the main 
sequence, but at much higher absolute luminosity than locally.

\begin{figure}[t]
 \includegraphics[width=\columnwidth]{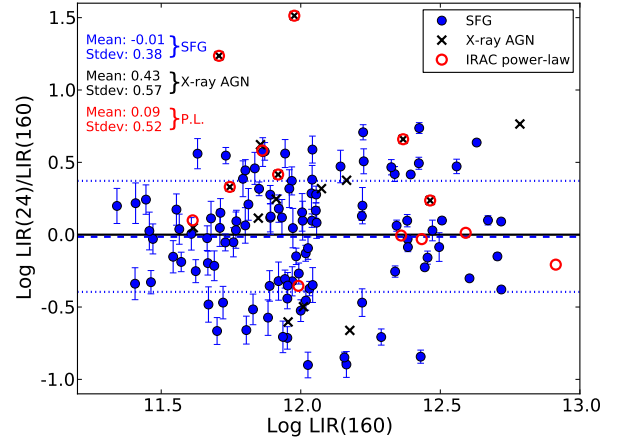}
 \caption{The scatter of LIR extrapolated for 1.5$<$z$<$2.5 sources from observed 24 $\mu$m using templates rescaled according to the sources offset from the main sequence, versus LIR measured from 160 $\mu$m. Blue circles are SFG, red circles are sources with power-law SED in the IRAC bands and black crosses are X-ray AGNs. The blue dashed and dotted lines indicate the overall mean and standard deviation for the SFG.}
 \label{fg:L24 vs L160 SSFR}
\end{figure}

For the practical application of deriving LIR for a galaxy of known redshift,
stellar mass, and rest frame 8~$\mu$m flux we can use Equation~\ref{eq:LCE01 vs SSFRoffset}
(or Equation~\ref{eq:nuLnu8/LIR vs SSFRoffset} in combination with template libraries other than CE01), 
stellar mass and the main sequence at the given redshift to derive a 
CE01 style template library as a function of LIR, which is then 
specific to this redshift and stellar mass (see Appendix~\ref{app:calibrate library}).
This library can then be applied
in the usual way to estimate LIR from the 24~$\mu$m flux and redshift.  
We have tested this method using it to estimate LIR(24) for the $1.5<z<2.5$ 
sources from the MIPS 24~$\mu$m fluxes as described.
The comparison with the LIR derived from the 160~$\mu$m photometry at the 
far-infrared peak is plotted in Figure~\ref{fg:L24 vs L160 SSFR}.
We can see from this figure that there is no bias in derived LIR(24) and 
no trends with LIR, albeit the $\lesssim$0.4~dex scatter is significant. 
We discuss this calibration for the use of 8~$\mu$m as a SFR indicator versus other calibrations in Sect.~\ref{sec:Discussion calibrations}.

%%%%%%%%%%%%%%%%%%%%%%%%%%%%%%%%%%%%%%%%%%%%%%%%%%%%%%%%%%%%%%%%%%
%
\section{IRS spectra of high-z galaxies} \label{sec:IRS spectra}
%
%%%%%%%%%%%%%%%%%%%%%%%%%%%%%%%%%%%%%%%%%%%%%%%%%%%%%%%%%%%%%%%%%%

\begin{figure}[t]
\begin{center}
 \includegraphics[width=1.05\columnwidth]{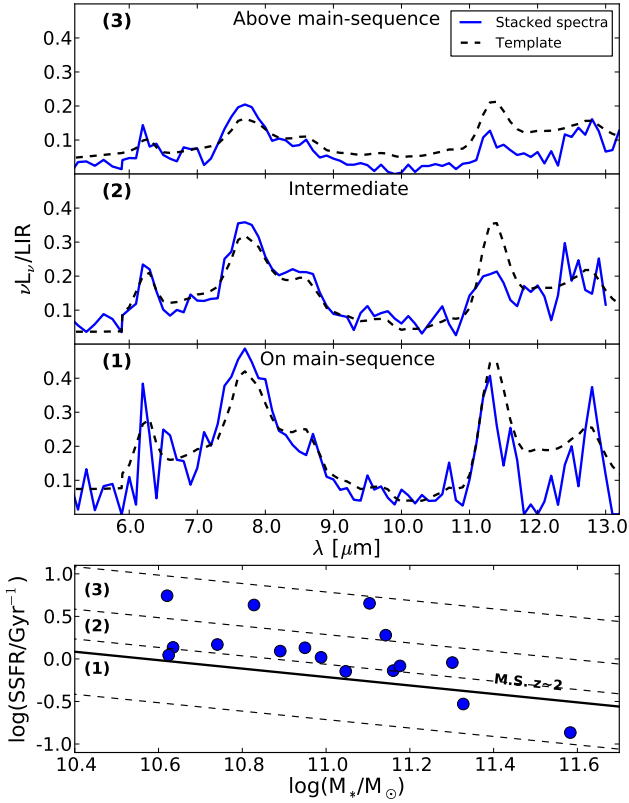}
 \caption{IRS spectra of z$\sim$2 galaxies and associted templates.
 {\it Top panels:}
 Template SEDs (black) scaled according to the SSFR distance from the main sequence, increasing from top to bottom, compared with stacked IRS spectra for non-AGN galaxies (blue). The three panels correspond to the selection bins as indicated by the number at the top-left of each panel. The templates are scaled according to the 160~$\mu$m flux and are not a fit to the spectra.
	  {\it Bottom:}
 The selection bins for the stacked spectra. The bins are at a constant SSFR distance from the main sequence, which is marked as a thick solid line.
 } 
\label{fg:IRS_SSFR_z2}
\end{center}
\end{figure}

\begin{figure}[t]
\begin{center}
 \includegraphics[width=1.0\columnwidth]{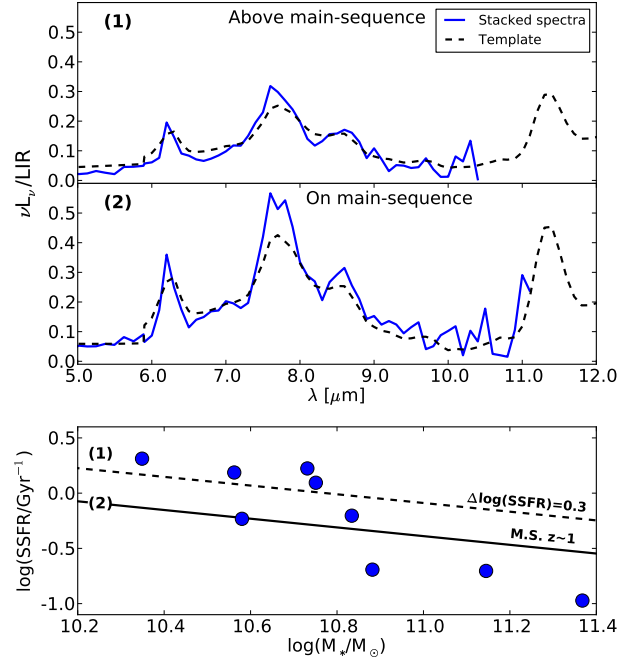}
 % IRS_MIPS_SSFR_z1.eps: 0x0 pixel, 300dpi, 0.00x0.00 cm, bb=13 30 598 761
 \caption{IRS spectra of z$\sim$1 galaxies and associted templates.
 {\it Top panels:}
 Template SEDs (black) scaled according to the SSFR distance from the main sequence, increasing from top to bottom, compared with stacked IRS spectra for non-AGN galaxies (blue). The two panels correspond to above and below the selection line indicated by the number at the top-left of each panel and illustrated in the bottom panel. The templates are scaled according to the 160~$\mu$m flux and are not a fit to the spectra.
	  {\it Bottom:}
 The selection bins for the stacked spectra. The sample is split at $\Delta$log(SSFR)$_{\rm MS}$=0.3 indicated by a dashed black line. The main sequence is marked as a thick solid line.
 }
 \label{fg:IRS_SSFR_z1}
\end{center}
\end{figure}

In the previous section we have derived a calibration for the SED 
templates as a function of the $\Delta$SSFR$_{\rm MS}$ distance from the 
main sequence.
The method in which we shift the filters across the spectral features by observing 
galaxies of slightly different redshifts 
(Figure~\ref{fg:nuLnuLir fit by SSFR}) can 
only provide limited information regarding the detailed spectra shapes,
which are consistent with the presence of PAHs.
In this section we extend this analysis using a sample of 24 
IRS spectra of z$\sim$2 log(LIR/L$_\odot$)$\approx$12 and 12 spectra of z$\sim$1 log(LIR/L$_\odot$)$\approx$11 sources in our field, which was first presented by 
\citet{Fadda10}.
In total, 31 of the IRS sources are detected by PACS at 160 $\mu$m and 6 of 
them have an associated X-ray detection and are flagged as AGNs.
The IRS spectra allow us to test the SED templates, which were assumed to have PAH features similar to those observed in local galaxies, but were calibrated 
above using photometry alone. 

We first focus on the 16 non-AGN z$\sim$2 objects. Due to the weak signal 
in each individual spectrum we stack several spectra 
in bins according to the distance of the galaxies from the main sequence.
The bottom panel in Figure~\ref{fg:IRS_SSFR_z2} shows the location of the galaxies on the SSFR versus stellar mass diagram and the three selection bins at constant distance from the main sequence.
Before stacking, the spectrum of each object is converted to the \nuLnu/LIR 
scale according to the individual LIR measured in the far-infrared by PACS. This avoids the few most luminous sources dominating the stacked result.
The spectra are then averaged in \nuLnu/LIR scale with equal weights. 
Equal weights prevent the few brightest and best S/N sources from dominating 
the final mean.
We then compare the stacked spectra to the modified templates as a function of
SSFR main sequence offset, as derived in Section~\ref{sec:by SSFR}. 
The templates are {\it not} fitted to the IRS spectrum - they are 
scaled by PACS 160 $\mu$m flux and extrapolate down to the mid-infrared around 
rest 6 $\mu$m. The result is plotted in Figure~\ref{fg:IRS_SSFR_z2}.

The IRS spectra are clearly PAH dominated. This strongly argues that
PAH emission rather than an AGN continuum must drive the `mid-infrared excess'
and associated mid-infrared based SFR overestimates, as noted by 
\citet{Fadda10} for this sample and \citet{Rigby08} and \citet{Murphy09} 
for other IRS samples. It is also consistent with the conclusions of \citet{Elbaz11} based on photometry.
In addition, the decrease in \nuLnu(8)/LIR when moving above the main 
sequence is indeed driven by a changing ratio of PAH to far-infrared, 
related to changing conditions in the interstellar medium.   
The stacked spectra of galaxies on the main sequence or slightly below it 
(region 1, bottom panel) are described very well by the rescaled template for 
the main sequence.
These galaxies with the strongest \nuLnu(8)/LIR relate to the ceiling value of 1~$\mu$m 
for the 6.2~$\mu$m PAH equivalent width reported in \citet{Fadda10}.

The templates were calibrated in Section~\ref{sec:by SSFR} 
according to photometry which covered mainly the 
7.7 $\mu$m PAH peak, but in the IRS spectrum the 6.2 and 11.3 $\mu$m peaks, as well as the 10 $\mu$m silicate absorption are also well matched. The z$\sim$2 main sequence galaxies indeed resemble 
scaled-up local templates even in their spectral details: PAH emissions, 
silicate absorption, continuum level.
This is in broad agreement with previous results that the SEDs of high 
redshift galaxies resemble scaled up SEDs of locals galaxies of lower 
luminosities (see the introduction). We here argue that this 
resemblance is based on the relation of the galaxies to the main sequence, 
rather than their absolute luminosities.

As we progressively move away from the main sequence (region 2 and 3 in Figure~\ref{fg:IRS_SSFR_z2}) the 
non-linearity in the mid- to far-infrared ratio is evident. Not only does the ratio of 
7.7 $\mu$m PAH to far-infrared decrease but the 6.2 and in particular 11.3 $\mu$m 
emissions are fainter relative to far-infrared as well.
The typical width of the main sequence at a given mass is 0.3--0.4~dex in SSFR \citep{Noeske07, Elbaz07} and so bin 3 can be considered fully above 
the main sequence and show the strongest effect.
While the templates are close to the spectra around 8~$\mu$m due to the way 
they were calibrated, they overestimate the emission from neutral PAHs at 
11.3~$\mu$m as well as the continuum level at wavelengths of 10~$\mu$m and 
above. These deviations compared to the stack of 4 weak IRS spectra  cannot 
be tested in the photometry because there are no 
filters that observe these rest-frame wavelengths at 1.5$<$z$<$2.5.

A similar picture arises from the sample of z$\sim$1 spectra. Nine galaxies with IR luminosity similar to local LIRGs and without an associated X-ray detection are detected by PACS 160~$\mu$m and 8 of them are detected at 100 $\mu$m as well. We split them at $\Delta$log(SSFR)$_{\rm MS}$=0.3 into two bins: on and above the main sequence. 
The stacked spectra are plotted in Figure~\ref{fg:IRS_SSFR_z1}.
The spectra are matched very well by the scaled templates and the relative difference in PAH strength between, on, and above main sequence is clearly visible.
The suppression of PAH strength is for this sample clearly associated 
with offset above the main sequence rather than higher LIR. Given the 
consistency of photometric and spectroscopic results, it is quantitatively
expressed by 
Figure~\ref{fg:CE01 rescale SSFR} and Equation~\ref{eq:nuLnu8/LIR vs SSFRoffset}.
Unfortunately at this redshift, the 11.3 $\mu$m PAH emission is outside the 
observed spectral range.

In a scenario where star forming galaxies above the main sequence are 
characterized by compact intense starbursts \citep[e.g.,][]{Wuyts11b, Elbaz11}, 
two factors can contribute to the weakening of all PAHs relative to the far-infrared for these galaxies.
First, high radiation field intensities in these compact regions can lead to inherently 
reduced PAH emission along with warmer large grain dust temperatures 
\citep[e.g.][]{DH01}.
Second, in the galaxies above the main sequence the PAH 
emission, even in the mid-infrared, could simply be obscured by higher column densities.
Given the mid-infrared extinction curve with its strong silicate feature around $\sim$10~$\mu$m,
such obscuration of a PAH
dominated spectrum will cause reduced 8.6 and 11.3~$\mu$m PAHs in
the wings of the silicate feature, compared to the 7.7$\mu$m one. 
This is indeed observed in some local galaxies \citep{Spoon00}. 
The S/N of the stack of four objects above the z$\sim$2 main sequence is not sufficient for such a test. 
From high S/N IRS spectra of local ULIRGs (thus above the main sequence) \citet{Veilleux09}
conclude that the obscuration of the PAH component is typically moderate: 
A$_V\lesssim$10. This analogy suggests intense radiation fields to be the
dominant cause of the PAH weakness, implying also warmer dust.

\citet{Kelson10} argued for a strong contribution of thermally pulsating asymptotic giant branch (TP-AGB)
stars to the
mid-infrared emission of galaxies from z=0 to z=2, that could in principle be
another cause for overestimated mid-infrared based star formation rates. Mid-infrared 
spectral evidence for AGB stars in integrated galaxy light indeed exists 
via detection of silicate emission in passively evolving local galaxies, with 
favorably low ratios of current to past star formation \citep{Bressan06}.
The IRS spectra of our star forming galaxies instead are dominated 
by luminous PAHs from active star formation, arguing against a noticeable
AGB contribution to their mid-infrared excess.  

We conclude that the 8~$\mu$m `excess' which caused the overestimation of 
SFR derived from MIPS 24~$\mu$m photometry for z$>$1.5 galaxies was caused 
almost entirely by underestimation of the PAH emission when using local templates.
The calibration derived in Section~\ref{sec:by SSFR}, that selects the templates by distance from the main sequence instead of LIR correctly predicts the 8~$\mu$m emission, which is dominated by the PAH emission.
The spectra of the SFG (no X-ray detection or IRAC-bands power-law) show no significant continuum contribution which could indicate an obscured AGN, though do not exclude the presence of a low luminosity one.

%%%%%%%%%%%%%%%%%%%%%%%%%%%%%%%%%%%%%%%%%%%%%%%%%%%%%%%%%%%
%
\section{The mid-to-far IR SED as a function of luminosity} \label{sec:by LIR}
%
%%%%%%%%%%%%%%%%%%%%%%%%%%%%%%%%%%%%%%%%%%%%%%%%%%%%%%%%%%%%

\begin{figure*}[t]
\begin{center}
  \includegraphics[width=0.99\textwidth]{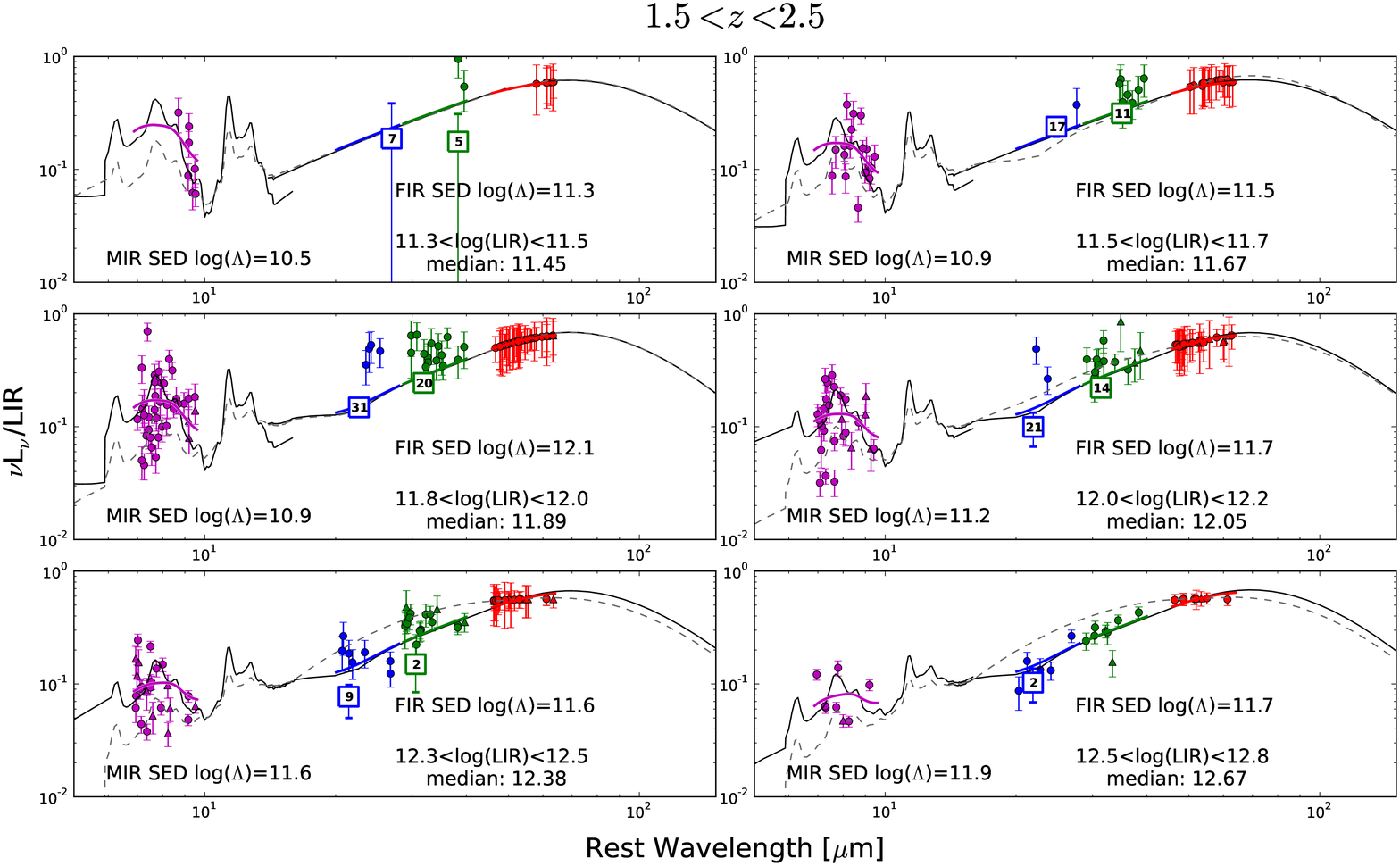} \\
  \includegraphics[width=0.99\textwidth]{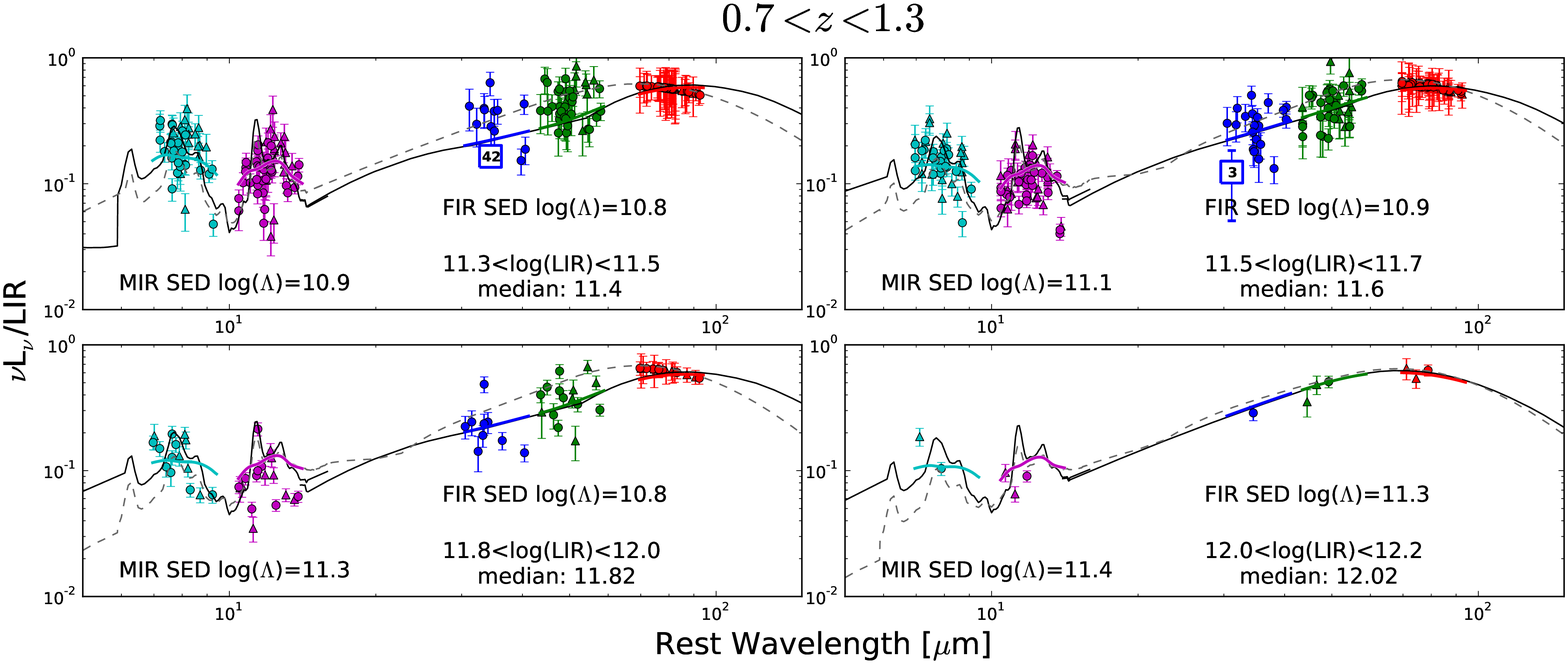}
  \caption{ {\it Top:} Rest-frame \nuLnu/LIR for various LIR bins of $1.5<z<2.5$ galaxies.
  Colors and symbols are identical to those used in Figure~\ref{fg:nuLnuLir fit by SSFR}.
  The original CE01 template that is associated with the bin 
  luminosity LIR is plotted as a dashed gray line.
  {\it Bottom:} Same as above for $0.7<z<1.3$ galaxies. The additional IRS 16~$\mu$m photometry is colored in cyan.}
  %\label{fg:CE01nuLnuLir_z2}
  \label{fg:nuLnuLir fit by LIR}  
\end{center}
\end{figure*}

\begin{figure}[t]
\begin{center}
 \includegraphics[width=1.05\columnwidth]{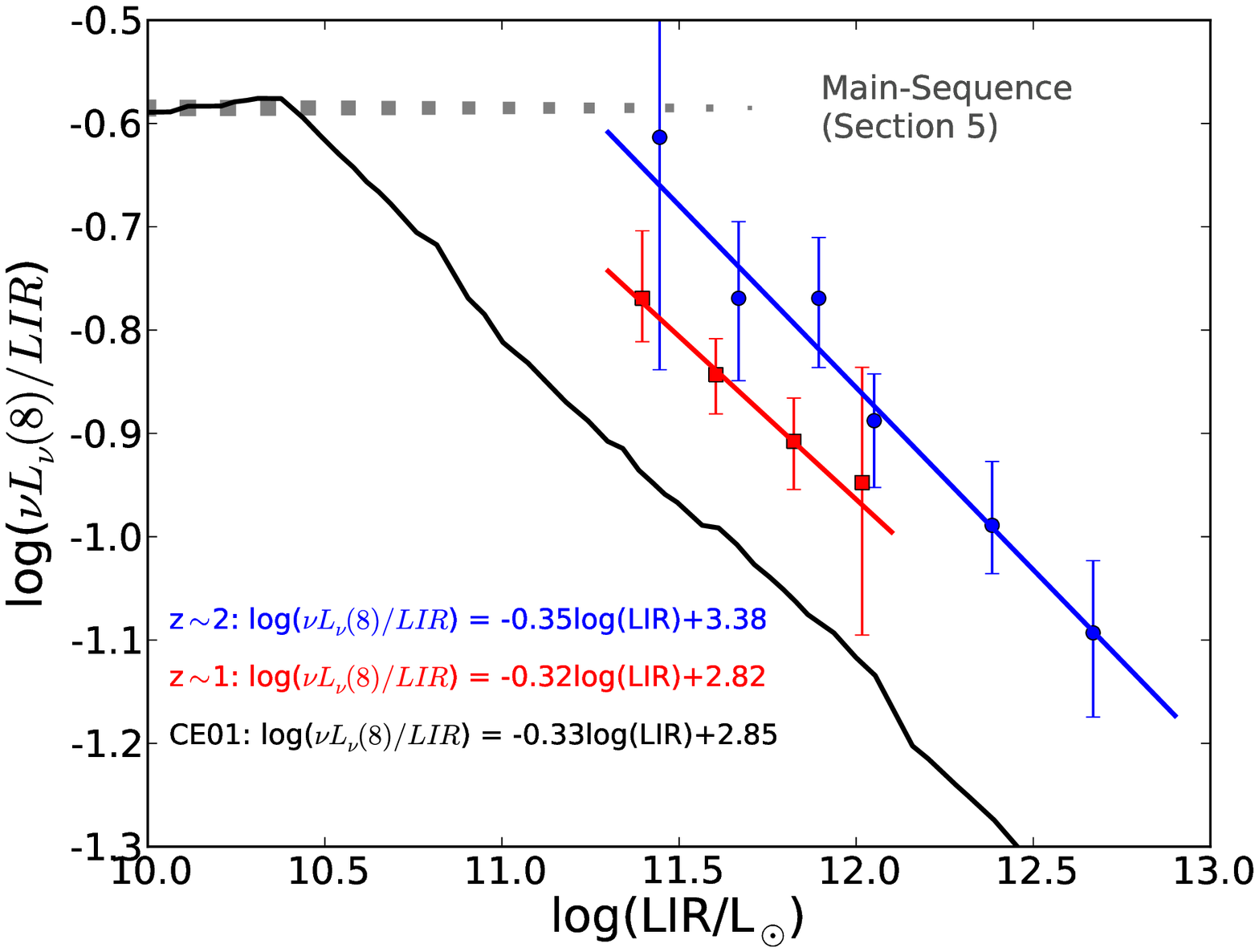}
 % CE01_rescale_curve_LIR.eps: 0x0 pixel, 300dpi, 0.00x0.00 cm, bb=13 175 598 616
 \caption{Mean log(\nuLnu(8)/LIR) vs. log(LIR) for z$\sim$2 (blue) and z$\sim$1 (red) galaxies. The relation for local galaxies as represented by the CE01 templates is plotted in black.
 The gray dotted line mark the main sequence \nuLnu(8)/LIR value as derived in Section~\ref{sec:by SSFR}.
 \nuLnu(8) value is the MIPS24 filter luminosity as would be measured for a galaxy at z=2.}
 \label{fg:nuLnu8/LIR rescale curve}
\end{center}
\end{figure}

In this section we will derive a benchmark by which SED libraries can be 
calibrated to match high redshift galaxies without required knowledge of 
the galaxy mass.
We use the methods described in section~\ref{sec:method} but in this case 
using narrow LIR bins, for the same z$\sim$2 and
z$\sim$1 parent samples.

The $1.5<z<2.5$ sample is divided into 6 LIR bins in the range 11.3$<$log(LIR/L$_\odot$)$<$12.8.
The $0.7<z<1.3$ sample is divided into similar LIR bins, however we are unable to populate the two highest 
luminosity bins over our fields due to the rarity of such objects at z$\sim$1.
Lower luminosities of the z$\sim$1 sample are limited by the depth of the 16 $\mu$m data and in order to maintain high 16~$\mu$m detection rate, we limit the lowest luminosity similar to that of the z$\sim$2 sample.

The result is plotted in Figure~\ref{fg:nuLnuLir fit by LIR} top and bottom for redshifts z$\sim$2 and z$\sim$1 respectively.
The LIR bin range is indicated in each panel.
For comparison, we also plot in dashed gray line the default CE01 template for the median luminosity in each bin. 
Figure~\ref{fg:nuLnu8/LIR rescale curve} summarizes the mid-infrared fit results by plotting for each LIR bin the mean \nuLnu(8)/LIR derived from the fitted template.
For reference, we add the \nuLnu(8)/LIR versus LIR for the local CE01 library 
templates. Again we specify for all redshifts \nuLnu(8)/LIR in terms of the 
z=2.0 MIPS 24 $\mu$m flux for the template.

There is a clear evolution with redshift of the relation between 
rest-frame \nuLnu(8~$\mu$m)/LIR and LIR. The local CE01 SEDs consistently 
underpredict the 8~$\mu$m emission at all redshift and luminosity bins.
This is the cause of the overestimation of SFR derived from 
24~$\mu$m at z$>$1.5 which was reported in previous studies, 
in comparison to local LIR--SED template relations.
At z$\sim$1 (Figure~\ref{fg:nuLnuLir fit by LIR}, bottom), the 24~$\mu$m filter 
observes rest-frame wavelengths of $\lambda > 10$~$\mu$m where the original 
CE01 SEDs and the newly fitted ones agree much better. This is because the 
region of the 11.3~$\mu$m PAH with its underlying continuum scales closer to 
linearly with LIR. Hence, the derived 24~$\mu$m based SFRs were accurate for z$<$1.5 galaxies. 
The rest frame 8~$\mu$m emission is underpredicted by the templates at 
z$\sim$1 as well, but it is not covered by the MIPS24 filter. A 16~$\mu$m 
filter is required to detect at z$\sim$1 an effect similar to the z$\sim$2
`mid-infrared excess'.

Interestingly, for our redshift bins the points in 
Figure~\ref{fg:nuLnu8/LIR rescale curve}  are nearly aligned and we pass a 
linear fit through them.
The slopes of the relations for z=2, 1 and 0 (the CE01 original library) are 
nearly identical. The curves for each redshift and log(L/L$_\odot$)$\sim$12 
are:
\begin{equation}
 \begin{array}{ll}
  \log({\nu}L_{\nu}(8)/LIR) = \\
  -0.35_{\pm 0.08} \log(LIR/10^{12}L_\odot) -0.86_{\pm 0.03} & :z\sim2 \\
  -0.32_{\pm 0.13} \log(LIR/10^{12}L_\odot) -0.96_{\pm 0.06} & :z\sim1 \\
  -0.33 \log(LIR/10^{12}L_\odot) -1.13 & :z\sim0
 \end{array}
 \label{eq:nuLnu8/LIR vs LIR}
\end{equation}

One can view this redshift evolution in two ways. The first is that the PAH 
strength \nuLnu(8)/LIR increases with redshift for a given LIR, i.e. 
evolution upwards in Figure~\ref{fg:nuLnu8/LIR rescale curve}.
The second is that high redshift galaxies have a similar
\nuLnu(8)/LIR as lower luminosity local galaxies, i.e. a shift to the right 
in Figure~\ref{fg:nuLnu8/LIR rescale curve} by $\sim$0.5 dex to z$\sim$1 and $\sim$0.8 dex to z$\sim$2.
The results of section~\ref{sec:by SSFR} favor the second description.
This in turn predicts that the relations described in Equation~\ref{eq:nuLnu8/LIR vs LIR} will flatten into the constant main sequence value of log(\nuLnu(8)/LIR)=-0.58 at lower luminosities, as also happens with local galaxies.
This description via a shift in the LIR associated with a given template 
to lower luminosities as the redshift increases is also in line with the 
results of previous work listed in the introduction.
Thus, the overestimation of 24 $\mu$m SFRs was not due to globally 
enhanced PAH strength at high redshifts, but due to 
a mismatch of template shapes with the associated LIRs in the library.
Same physical conditions in galaxies of a given LIR at all redshifts cannot be assumed.
We will discuss the relation between the above description (SEDs depend on LIR) and 
our preferred description from Section~\ref{sec:by SSFR} (SEDs depend on distance from the main sequence)
in Section~\ref{sec:Discussion}.

Adapting the CE01 library for use with high redshift galaxies is a simple matter of assigning new luminosities to each template. For example, from Equation~\ref{eq:nuLnu8/LIR vs LIR} we get that a CE01 template, originally associate with with luminosity $\Lambda_{CE01}$, should be rescaled to a new luminosity $L_{z=2}$ with which it is associated at z$\sim$2, using the relation:
\begin{equation}
 \log(L_{z=2}) = 0.943 \log(\Lambda_{CE01}) +1.51
\label{eq:CE01_template_rescale_z2}
\end{equation}
We stress that the negative slopes in Equation~\ref{eq:nuLnu8/LIR vs LIR} are 
valid only in the luminosity range constrained by our PACS data. 
Locally, $\log({\nu}L_{\nu}(8)/LIR)$ 
levels off at $\sim$-0.58~dex for low luminosity objects \citep{CE01} as illustrated in Fig~\ref{fg:nuLnu8/LIR rescale curve}. 
Since low luminosities are not sufficiently constrained in our z$\geq$1 data (16 $\mu$m limited) or z$\geq$2 data (160 $\mu$m limited),
it appears prudent to either not apply Equation~\ref{eq:nuLnu8/LIR vs LIR} 
below the luminosities that are constrained at a given redshift or at least 
assume a similar leveling off at $\log({\nu}L_{\nu}(8)/LIR)=-0.58$, indicated by the fading dotted line in Fig~\ref{fg:nuLnu8/LIR rescale curve}.
We compare the template calibration derived in this section with other calibrations in Sec~\ref{sec:Discussion calibrations}.

The far-infrared range covered by PACS, i.e. rest-frame wavelengths of 20--90~$\mu$m in Figure \ref{fg:nuLnuLir fit by LIR}, shows a behavior consistent with 
the scaling of the SEDs in the mid-infrared.
The best fitting templates for the z$\sim$2 sample are in the range of $11.3<\log(\Lambda_{CE01})<12.1$. In practice, these CE01 templates have very 
similar \nuLnu($\lambda$)/LIR across the rest-frame wavelengths probed by 
the PACS filters and the exact $\Lambda_{CE01}$ is degenerate in this luminoity and wavelength range.
At the highest luminosities (bottom panels), local galaxies would 
typically have a hotter ($T_{dust}\sim50$~K) ULIRG-like rest frame 
far-infrared SED, plotted in dashed line. The observed SEDs resemble 
instead LIRG-like (log(LIR/L$_\odot$)$>$11) SED shapes with their luminosity elevated by $\sim$0.4--0.8 dex \citep[see also][]{Muzzin10, Rex10}.
At the lower luminosities, the 70 and 100 $\mu$m filters only detect the upper part of the population scatter. However, when taking into account the stacks (in some cases containing much larger number of sources than the detections) the mean fits the local LIRG templates.
At the highest luminosities we detect nearly the full sample at all wavelengths and the match to the best fit template is quite striking - almost as if we are sampling the same galaxy at different 1.5$<$z$<$2.5 redshifts.

%%%%%%%%%%%%%%%%%%%%%%%%%%%%%%%%%%%%%%%%%%
%
\section{AGNs vs. SFG} \label{sec:AGNs}
%
%%%%%%%%%%%%%%%%%%%%%%%%%%%%%%%%%%%%%%%%%%

\begin{figure}[t]
\begin{center}
 \includegraphics[width=\columnwidth]{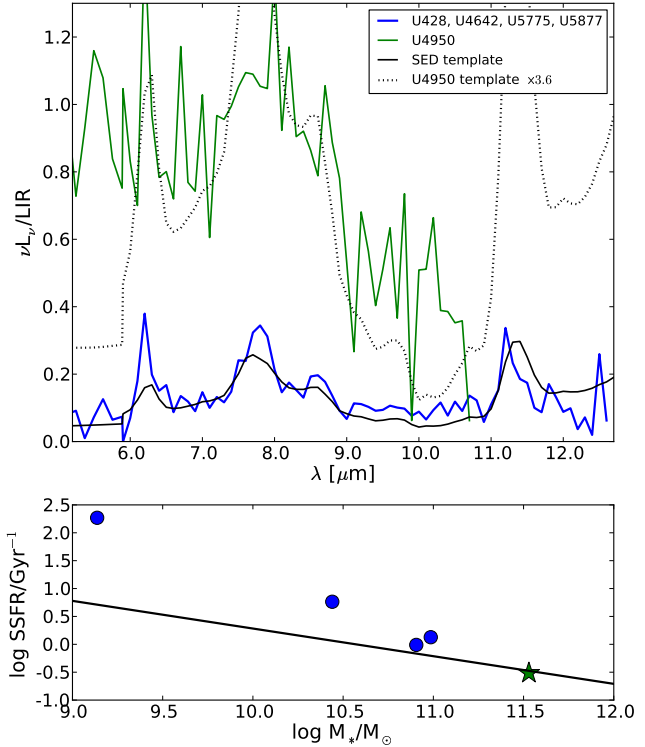}
 \caption{{\it Top:} Stacked IRS spectrum of four z$\sim$2 AGNs (blue). Spectrum of another AGN (U4950) is plotted    in green.
 The predicted stacked spectrum from the rescaled CE01 templates is in solid black. The templates are fitted to the PACS 160~$\mu$m flux, not to the spectrum.
 {\it Bottom:} The location of the AGN hosts with respect to the main sequence. U4950 is plotted as a green star.
 }
 \label{fg:IRS_AGNs}
\end{center}
\end{figure}

In all our samples thus far we have removed all sources detected in the {\it Chandra}~2~Ms surveys in GOODS-N and GOODS-S.
The stacked IRS spectra of the SFGs (Section~\ref{sec:IRS spectra}) clearly show mid-infrared emission dominated by PAHs typical of star formation and no significant emission from obscured AGNs.
We now turn to inspect the mid-infrared emission of the X-ray sources. At redshifts of 2 and with the depth of the 2~Ms catalogs, it is safe to assume that all the X-ray sources detected by PACS 160 $\mu$m are AGN hosts. Only a small fraction of the highest LIR sources are suspected to have their X-ray emission dominated by star formation (see Section~\ref{sec:Data}), which does not rule out the possible presence of an AGN.

In Section~\ref{sec:by SSFR} Figure~\ref{fg:L24 vs L160 SSFR}
we compared the LIR as derived from 24~$\mu$m to the LIR derived from 160~$\mu$m. In the figure we also plotted the X-ray AGNs of our sample (black x marks) and sources that show a clear power-law SED in the 3.6--8.0 $\mu$m {\it Spitzer}-IRAC bands (red circles). 
If AGN-heated dust contributed significantly to the mid-infrared emission, we should expect an enhanced L24/L160 for these galaxies, due the mid-infrared continuum emission from warm circumnuclear dust. This would lead to wrong extrapolation from mid- to the far-infrared and overprediction of the luminosity.
In the figure, most AGNs are well within the scatter of the normal SFGs. A small bias does exist and AGN hosts show a small statistical enhancement in $\sim$8~$\mu$m flux. In only 2 out of 18 AGN-hosting galaxies the mid-infrared emission is significantly enhanced compared to the typical scatter for the SFG in the sample.

The \citet{Fadda10} IRS sample includes 5 z$\sim$2, X-ray AGNs which are detected with PACS 160~$\mu$m. These were excluded from the analysis in Section~\ref{sec:IRS spectra}.
The individual spectra are quite noisy. However, when normalizing by each galaxy's LIR, four of the spectra overlap within the noise, while one (U4950) shows significantly brighter mid-infrared emission.
In Figure~\ref{fg:IRS_AGNs} we plot the stacked spectrum of the four z$\sim$2 X-ray detected sources from the \citet{Fadda10} sample. We use the same stacking (spectral averaging) procedure as in Sec~\ref{sec:IRS spectra}.
The spectrum of U4950 is plotted separately.
Overplotted is the expected mean spectrum from the CE01 templates, rescaled according to the results from Section~\ref{sec:by SSFR} and stacked in a similar way as the spectra. The templates which were scaled according to the non-AGNs also match the X-ray sources quite well. In particular, there is no enhanced continuum level which may be attributed to AGN-related circumnuclear dust emission.
The stacked AGN mid-infrared spectrum shows all the typical features of a SFG and the relative AGN contribution must be low, as also concluded by \citet{Fadda10}.

The spectrum of U4950 is significantly different and hence was not included in the stack. In addition to having much higher rest mid-infrared fluxes than expected from a SFG, which as judged from the LIR lies exactly on the \citet{Rodighiero10} main sequence (Figure~\ref{fg:IRS_AGNs} bottom panel), it seems to be flatter and elevated by an added continuum rather than by an enhanced PAH emission.
The SFG template expected for U4950 according to its mass and LIR is plotted as a dotted line in Figure~\ref{fg:IRS_AGNs}, scaled up by a factor 3.6 to match its observed 24~$\mu$m flux. The U4950 spectrum is fundamentally different from such a scaled SFG template and is dominated by AGN continuum which is evident shortward of 6 $\mu$m, where AGN continua can be cleanly identified \citep{Laurent00}.

This behavior is consistent with the X-ray properties.
We convert from $L_x$(2--10~keV) to bolometric luminosity $L_{\rm AGN}$ using the relation of \citet{Maiolino07} and a factor 7 \citep{Netzer07} conversion from optical to bolometric luminosity: 
\begin{equation}
 \log(L_{\rm AGN}) = \frac{\log(L_x)-11.78}{0.721} + \log(7)
 \label{eq:Lx to LAGN}
\end{equation}
The IR, X-ray, and AGN bolometric luminosities are summarized in Table~\ref{tab:IRS AGNs}.
U4950 has a significantly higher log(L$_{\rm AGN}$/LIR)
than the next highest ratio by more than 0.8~dex and more than 2~dex higher than the rest of the AGNs in this sample. 

\citet{Bauer10} derived the following relation between mid-infrared and X-ray AGN luminosity:
\begin{equation}
 \log \left( \frac{L_{5.8\,{\rm \mu m}}}{\rm erg\,s^{-1}} \right) = 1.21\pm0.06 \cdot \log \left( \frac{L_{2-10 {\rm keV}}}{\rm erg\,s^{-1}} \right) -8.7\pm2.6
 \label{eq:Bauer10}
\end{equation}
This relation has been derived for much more luminous AGNs than the ones in our sample, but it is also consistent with the results of \citet{Lutz04} which were derived for low luminosity AGNs.
Combining this relation with equation~\ref{eq:nuLnu8/LIR vs LIR} for z$\sim$2
and assuming that the AGN continuum is roughly flat in ${\nu}F_\nu$ between 5.8 and 8 $\mu$m,
 we get the condition that in order for a z$\sim$2 AGN to dominate the 
emission close to rest frame 8~$\mu$m, the relation between the X-ray luminosity and the LIR must satisfy:
\begin{equation}
 \log \left( \frac{L_{2-10\, {\rm keV}}}{\rm erg\,s^{-1}} \right) \gtrsim \left\{
  \begin{array}{ll}
   0.537 \log\left( \frac{LIR}{\rm erg\,s^{-1}} \right) + 19.71  &  \log\left( \frac{LIR}{L_\odot} \right) \geq 11.3 \\
   0.826 \log\left( \frac{LIR}{\rm erg\,s^{-1}} \right) + 6.71  &  \log\left( \frac{LIR}{L_\odot} \right) < 11.3
  \end{array} \right.
\end{equation}
Or in terms of bolometric AGN luminosity (equation~\ref{eq:Lx to LAGN}):
\begin{equation}
 \log \left( \frac{L_{\rm AGN}}{\rm erg\,s^{-1}} \right) \gtrsim \left\{
  \begin{array}{ll}
   0.745 \log\left( \frac{LIR}{\rm erg\,s^{-1}} \right) + 11.84  &  \log\left( \frac{LIR}{L_\odot} \right) \geq 11.3 \\
   1.145 \log\left( \frac{LIR}{\rm erg\,s^{-1}} \right) - 6.18  &  \log\left( \frac{LIR}{L_\odot} \right) < 11.3
  \end{array} \right.
\end{equation}
The condition for log(LIR/L$_\odot$)$<$11.3 assumes that \nuLnu(8)/LIR=-0.58 below this luminosity, as suggested in Section~\ref{sec:by LIR}.
A similar, albeit more complicated relation that involves $\Delta$log(SSFR)$_{\rm MS}$ can be derived by combining Equation~\ref{eq:Bauer10} with Equation~\ref{eq:nuLnu8/LIR vs SSFRoffset}.

Overall, for our sample where log(LIR/L$_\odot$)$\sim$12, an AGN bolometric 
luminosity of log(L$_{\rm AGN}/{\rm erg~s}^{-1}$)$>$45.8 or X-ray luminosity
log(L$_{2-10\,{\rm keV}}/{\rm erg~s}^{-1}$)$>$44.2 is required to dominate 
the 8 $\mu$m emission.
Only U4950 reaches these AGN luminosities in the IRS sample used here and indeed shows a clear excess in its 24 $\mu$m flux. 
In general, such high AGN luminosities are rare in the GOODS fields which explains why most of our (far-infrared bright) AGNs are indistinguishable from the SFGs in their mid-to-far IR emission \citep[see also][]{Nordon10, Elbaz10}.
The changed slope at log(LIR/L$_\odot$)$<$11.3 corresponds to a 
population that is dominated by galaxies on or below the main sequence. Here, 
the 8~$\mu$m luminosity associated with the star formation decreases 
proportionally to LIR (constant \nuLnu(8)/LIR), while above that threshold the 8$\mu$m luminosity will 
vary more slowly with LIR due to the changing mid- to far-infrared SEDs.
The ratio of AGN luminosity and LIR
at which the AGN dominates the 8 $\mu$m emission hence clearly depends on 
sample selection.

\citet{Fadda10} fitted and subtracted the PAH emission from the stacked spectra of all the IRS sources for which the mid-infrared spectra do not appear to be dominated by an AGN.
They were thus able to constrain the mean AGN contribution to the 6 $\mu$m continuum and derived a mean intrinsic AGN L$_{2-10 {\rm keV}} = 1\times 10^{42}$ erg~s$^{-1}$.
According to the above analysis, such luminosities are nearly two orders of magnitude lower than the AGN luminosities required in order to create a true broadband 24~$\mu$m `excess' in massive ($\sim10^{11}$~M$_\odot$), z$\sim$2, main sequence galaxies.
Given that on average AGN hosts seem to have similar SFR as non-active 
massive galaxies 
at same redshift \citep{shao10}, this will likely still be true for many less 
far-infrared luminous objects, but a higher fraction of quenched sources with 
24~$\mu$m emission dominated by AGNs may be expected.

\begin{table}[t]
\begin{center}
\caption{\label{tab:IRS AGNs} IRS sources from the \citet{Fadda10} sample with an X-ray detection.}
\begin{tabular}{@{}lllll}
\hline
ID & log L$_{2-10\,{\rm keV}}$\footnote[1]{Fadda et al. (2010)} & log L$_{\rm AGN}$\footnote[2]{Equation \ref{eq:Lx to LAGN}} & log LIR\footnote[3]{This work}& log \nuLnu(MIPS24)$^{c}$ \\
   & [erg s$^{-1}$] & [erg s$^{-1}$] & [L$_\odot$] & [L$_\odot$] \\
\hline
U428  & $<$42.43 & $<$43.36 & 12.07& 11.40\\
U4642 & 42.54  & 43.51      & 11.86& 11.42\\
U4950 & 44.28  & 45.91      & 11.98& 11.93\\
U5775 & $<$42.54 & $<$43.51 & 12.16& 11.28\\
U5877 & 43.96  & 45.47      & 12.37& 11.47\\
\hline
\end{tabular}
\end{center}
\end{table}

%%%%%%%%%%%%%%%%%%%%%%%%%%%%%%%%%%%%%%
%
\section{Discussion} \label{sec:Discussion}
%
%%%%%%%%%%%%%%%%%%%%%%%%%%%%%%%%%%%%%%
\subsection{The infrared SED reflects ISM conditions rather than total LIR}

Our findings from the previous sections can be summarized as follows: 
The SED shape and in particular \nuLnu(8)/LIR correlate best with the 
offset from a sloped reference line in the SSFR(M$_*$) diagram.
Over our limited mass range, the dependence on the exact main sequence 
slope is weak -- it is 
the normalization that is more important. The redshift evolution of SEDs 
in this picture is driven by the shift of the main sequence to higher SSFR 
at higher redshift. The SEDs also correlate (slightly worse and with 
explicit redshift dependence) with the absolute LIR, and the redshift 
evolution is manifested here as a shift in the LIR associated with each 
template shape.

If infrared SED shape is 
tied to a typical local radiation field intensity \citep[e.g.][]{DH01},
which in turn is related to the efficiency SFR/M$_{\rm Gas}$ (SFE)
by which molecular gas is converted to stars (the number of young 
stars per gas mass available to irradiate it),
then it will also be linked to the SSFR, since 
${\rm SFE} = {\rm SSFR} \cdot M_* / M_{\rm Gas}$.
Because both gas fraction M$_{\rm Gas}$/(M$_{\rm Gas}$+M$_*$) 
and main sequence SSFR rise towards higher redshift, the increase in gas 
fraction and the increase in SSFR will partly compensate and SFE will be 
more closely linked to the SSFR offset from the main sequence.
SFR/M$_{\rm Gas}$ is lower (by factors 4--10) in z$\sim$2 high redshift main sequence galaxies,
compared to local ULIRGs with similar SFR but lower 
total gas fractions \citep[e.g., ][]{Genzel10}.
At z$\sim$0, radiation field intensity, SFR/M$_{\rm Gas}$ and SSFR offset 
are empirically correlated with LIR, largely driven by the peculiar properties
of merger driven (U)LIRGs and closely linked to their compactness, as also
reflected in the local compactness/temperature relation \citep{Chanial07}.
The larger gas fractions at higher redshifts permit larger LIR
before invoking special events like mergers and will shift correlations with
LIR.
The preferred connection between SED shape and SSFR offset from the main 
sequence reflects that the infrared SED shape is physically linked to 
the local ISM conditions, and only indirectly to LIR. In line with
these arguments, sizes of equivalently luminous IR galaxies change to higher 
redshift, with implications on their SED \citep{Rujopakarn11a, Rujopakarn11b}.

Far infrared fine-structure emission lines and in particular [CII] exhibit 
a similar behavior to the PAHs. Compact luminous objects with intense 
radiation fields such as local ULIRGs show a [CII] deficit relative to 
the far-infrared continuum emission \citep{Malhotra97, Malhotra01, 
Contursi02, Luhman03} with arguments for a direct physical link to PAH 
\citep{Helou01}.
The [CII] line with a wavelength of 158 $\mu$m is much less likely to be 
attenuated than the mid-infrared PAHs and the deficit is intimately related 
to the radiation fields around the star forming regions and the resulting 
structure of the HII and photo-dissociation regions (PDRs).
While there are no [CII] observations for our specific targets, the 
qualitative findings for [CII] both locally and at high redshift are 
consistent with the behavior of \nuLnu(8)/LIR.
There is a high-z [CII] deficit - compared to local it is setting in at 
higher LIR, but at similar ratio of LIR and gas mass (SFE), equivalent to 
similar main sequence offset \citep{Gracia11}. This is directly analogous to 
our finding
of a relation \nuLnu(8)/LIR to LIR that evolves with redshift, but a single
relation with main sequence offset.
The connection between SFE and offset from the main sequence which was 
described above, together with the analogy between the [CII] deficit and 
decreasing \nuLnu(8)/LIR in galaxies above the main sequence, suggest that 
the two observed phenomena are likely related and intense radiation fields to be 
the dominant cause of the PAH weakness.

To zeroth order, the redshift dependent main sequence can be seen as a 
reference and with the distance from this reference a number of observables change together: 
The SSFR (by definition), 
the star formation efficiency SFR/M$_{\rm Gas}$ \citep{Genzel10}, 
morphology and compactness \citep{Wuyts11b, Elbaz11}
the mid- to far-infrared SED \citep[this work, ][]{Elbaz11}, the
far-infrared SED shape tracing the large grain temperature \citep{Elbaz11}, 
and far-infrared fine-structure emission lines deficits \citep{Gracia11}.
Compactness due to mergers is likely a key factor in changing the local
conditions in the ISM that drive the scaling relations for these observables. 
The link of low \nuLnu(8)/LIR mid- to far-infrared SED shape with 
spatial compactness, previously indicated from local universe evidence 
mentioned in the introduction, is strongly supported both at low
and at high redshift in the analysis of \citet{Elbaz11}. 
 
\subsection{A simple model connecting cloud conditions and global SED}

Galaxy-integrated LIR is not a fundamental parameter that determines the 
conditions in the molecular clouds and PDRs which produce the IR emission. 
Rather, it has an empirical correlation with these conditions. 
Borrowing the terms from thermodynamics, a clear distinction should be kept between 
total {\it extensive} quantities which are summed over the entire
galaxy (LIR, stellar and gas masses \ldots), and 
{\it intensive} quantities which are 
averaged, such as SSFR, SED shape, gas depletion time scale \ldots .
It is easy to imagine a galaxy in which everything is doubled 
(every cloud and star becomes two etc.): extensive quantities like total 
LIR will be doubled, but intensive ones such as SED shape and SSFR will remain 
unchanged. The relations between extensive and intensive quantities will 
change in such a scaling. The SED shape as an intensive quantity that 
is related to local cloud physics should more
directly scale with another intensive quantity such as the SSFR or its offset
from the main sequence.   
The specific star formation rate {\it at a given redshift}, which is 
proportional to LIR/M$_*$, offers a natural scaling reference: LIR and M$_*$ 
are measurable and their ratio does not change when scaling as described.

We adopt the simplified hypothesis that main sequence galaxies at all of our 
redshifts form their stars in a single type of 
star forming molecular clouds (which for brevity, we will simply refer to as clouds)
and discuss the implications for our findings as well as other scaling relations 
below in a toy model.
Main sequence galaxies with different numbers of these clouds will thus 
have similar mid-to-far IR SED but a different LIR. Galaxies above the main 
sequence will have different local cloud properties.
 
If the number of the clouds were proportional to the stellar mass, more stars 
would mean proportionally more of the same clouds and LIR emission. Our 
adopted \citet{Rodighiero10} main sequence argues instead that the number of 
clouds scales less than proportional with the stellar mass when moving along
the main sequence.  This agrees
with the finding in  Section~\ref{sec:iso nuLnu8/LIR} that galaxies with 
constant \nuLnu(8)/LIR are not at constant SSFR, but have a SSFR(M$_*$) 
dependency. 

Comparing main sequence galaxies of similar mass at different redshifts in this picture, 
we expect more of the same clouds per stellar mass at higher redshift.
If similar stellar masses imply roughly similar galactic radii, then the 
high-z galaxies have 
similar molecular clouds with less empty volume between them, meaning a higher surface brightness.
Higher surface brightness of main sequence galaxies is indeed observed by \citet{Wuyts11b}.

We explore the scenario of a single type of star forming cloud in a toy model.
One can define a dimensionless efficiency parameter for star formation as:
\begin{equation}
 \frac{\dot{\rho}_*}{\rho_{\rm Gas}} \tau_{ff} = \epsilon
\end{equation} 
where $\dot{\rho}_*$ is the SFR per volume element, $\rho_{\rm Gas}$ is the gas density and $\tau_{ff}$ is the gas free fall time scale ($\tau_{ff} \propto \rho^{-1/2}$) of a gravitationally bound cloud.
$\epsilon$ is determined by the micro-physics inside the molecular cloud and we will assume it to be a global constant.
This translates to the {\it local} Kennicutt-Schmidt (KS) relation 
\citep{Schmidt59, Kennicutt89, Kennicutt98b}:
\begin{equation}
 \dot{\rho}_* \propto \epsilon {\rho_c}^{\alpha}
 \label{eq:local KS law}
\end{equation}
where we assume $\dot{\rho}_*$ and $\rho_c$ to be averages over the star forming
cloud (as opposed to the entire galaxy volume), indicated by the subscript c.
Traditionally $\alpha = 3/2$, however due to various 
measurements deriving values in the range of 
$1<\alpha<1.7$ \citep{Bouche07,Kennicutt07,Bigiel08, Genzel10} we leave 
this as a general parameter. The free-fall timescale is only relevant for 
a gravitationally bound cloud set to collapse and form stars, not for the 
averaged density over the entire galaxy.
Hence, the above is a local relation applicable to a small-scale molecular cloud entity.
We would like to integrate over
the galaxy volume to derive the total, extensive measurables: SFR and mass.
To do this, we define a filling factor $f$ that describes the fraction of 
the total volume $V$ of the galaxy which is occupied by molecular star 
forming clouds.The integrated SFR over the volume $V$ is then:
\begin{equation}
 \dot{M}_* \propto \epsilon \rho_c^\alpha f V
\end{equation}
We can now divide by the stellar mass $M_*$ and use $\rho_c f V = (M_{\rm Gas}/M_*) M_*$ to eliminate the volume and get:
\begin{equation}
 \frac{\dot{M}_*}{M_*} \propto \left( \frac{M_{\rm Gas}}{M_*} \right) \left(\epsilon \rho_c^{\alpha-1} \right)
\label{eq:SFR/m vs. rho}
\end{equation}
where M$_{\rm Gas}$ is the galaxy integrated molecular gas mass.

The right-most term in parenthesis, that includes the average {\it cloud} density
and the efficiency parameter, describes the local conditions in the star forming regions 
and hence is associated with the IR SED shape.
All main sequence galaxies will have 
the same value in these parentheses, and also the same SED. Galaxies above the 
main sequence will boost the ${\dot{M}_*}/{M_*}$ by changed cloud 
conditions, going along with SED and other changes.

From Equation~\ref{eq:SFR/m vs. rho}, galaxies with the same IR SEDs will tend to 
have similar SSFR, modified by their gas fractions.
On a SSFR versus M$_*$ diagram, such constant-SED galaxies (indicated by 
constant \nuLnu(8)/LIR in this study) will lie on a slope which follows 
the slope of the M$_{\rm Gas}$/M$_*$ relation. Comparing different redshifts,
changing gas fractions \citep{Tacconi10} would mediate 
in Equation~\ref{eq:SFR/m vs. rho} the change of main sequence SSFR with redshift.
This change is needed to have same cloud conditions and SED at different 
SSFR, but at the same SSFR offset 
from the redshift-dependent main sequence $\Delta$log(SSFR)$_{\rm MS}$.

In local galaxies, molecular gas fractions decrease gently with stellar mass, 
though with a large scatter \citep{Saintonge11a}. The slope of the local 
main sequence (in SSFR versus M$_*$) is very similar \citep{Brinchmann04, 
Salim07, Peng10}, in full agreement with the toy model.
At higher redshifts, the dependency of M$_{\rm Gas}$/M$_*$ on M$_*$ is yet somewhat unclear.
\citet{Tacconi10} measured the molecular gas fraction in a sample 
of z$\sim$1 and z$\sim$2 galaxies lying close to their respective main sequence. 
Their sample size and spread in M$_*$ is not yet sufficient to conclusively  
probe for trends of the gas mass fraction with stellar mass. Sample size
also limits the conclusions of \citet{Daddi10a} who find a near constant 
M$_{\rm Gas}$/M$_*$ for six z$\sim$1.5 main sequence galaxies.

Another way to look at equation~\ref{eq:SFR/m vs. rho} is to divide both 
sides by the gas to stars mass ratio to get a dependency on the gas depletion timescale $\tau_{dep} = M_{\rm Gas}/\dot{M}_*$ instead of SSFR:
\begin{equation}
 \tau_{dep}^{-1} \propto \epsilon {\rho_c}^{\alpha-1}
 \label{eq:tau_dep}
\end{equation}
$\tau_{dep}$ is an intensive quantity that is directly related to the 
process of star formation, unlike SSFR which involves the mass of older stars.
For local galaxies, \citet{Saintonge11b} find a relatively tight correlation between $\tau_{dep}$ and the SSFR across two orders of magnitude in SSFR: $\log(\tau_{dep}) = -0.724 \log({\rm SSFR}) +1.54$.
Interestingly, when they scale the SSFR by the change of the main sequence with redshift, the relation agrees with z$\sim$1 and z$\sim$2 SFGs, though the statistics are low.
In Equation~\ref{eq:tau_dep}, the measureable quantity $\tau_{dep}$ is equivallent to local cloud conditions and hence closely and directly related to the resulting IR SED.
If indeed $\tau_{dep}$ corresponds to the resulting IR SED shape, then this supports our conclusion from Section~\ref{sec:by SSFR} that the relation between SSFR and SED shape scales with redshift like the main sequence.
At a given redshift, the decrease of $\tau_{dep}$ with increasing SSFR represents the change in molecular cloud properties as galaxies move away from the main sequence towards `bursty' compact sources. 

The above suggests that the main sequence up to z$\sim$2 is composed of galaxies with rather uniform molecular cloud properties.
The difference in integrated quantities such as LIR is mostly due to the total number of clouds.
For normal galaxies, dust optical depths are low at the mid- to far-infrared 
wavelengths studied 
here. Observables will thus be less sensitive to second order effects of 
arrangement of these clouds than shorter wavelengths.
As we gradually move away from the main sequence to higher SSFR the properties of the clouds gradually change and $(\epsilon {\rho_c}^{\alpha-1})$ increases, meaning $\tau_{dep}$ decreases and the \nuLnu(8)/LIR decreases, representing denser star forming regions. In this toy model we cannot differentiate between an increase of the average gas density of the clouds $\rho_c$ or the efficiency $\epsilon$.

The global KS law $\Sigma_{SFR} \propto \Sigma_{\rm Gas}^\alpha$ uses the integrated 
molecular gas-mass surface density (total M$_{\rm Gas}$ over area) and integrated 
SFR surface density.
When considering only main sequence galaxies or galaxies parallel to the main sequence in general,
the power expected to be retrieved is $\alpha \approx 1$ instead of the theoretical $\alpha=1.5$ or in general the $\alpha$ of the local KS law (Equation~\ref{eq:local KS law}),
which could partially explain low $\alpha$ values quoted in recent works that distinct between types of galaxies and atomic versus molecular gas masses \citep[e.g., ][]{Gao04, Bigiel08, Leroy08, Genzel10}.
This is because according to the above suggested model and along lines of constant $\Delta$log(SSFR)$_{\rm MS}$, 
the increase in the surface densities is due to filling of the empty volumes by more of the same molecular clouds, increasing mean surface densities for SFR and for (molecular) gas proportionally.
$\alpha=1.5$ may still apply to the individual, or even fragments of molecular clouds (local law), if it were possible to measure it on small enough scales.
The various measured values for the power in the KS law highly depend on the definitions, methods and selections used. For a discussion on the various measurements of the KS slope see e.g., \citet{Genzel10}.

How does equation~\ref{eq:SFR/m vs. rho} relate to the correlation of SED 
shapes with LIR? This is the original way in which the CE01 library has been 
constructed for local galaxies and the correlation is also seen in high-z 
galaxies (Section~\ref{sec:by LIR}, Figure~\ref{fg:nuLnu8/LIR rescale curve}).
As an extensive quantity, LIR must be correlated with the clouds that emit the SED through other parameters or circumstance.
For example, locally, luminosities of LIR$>10^{12}$~L$_\odot$ require either a very massive main sequence galaxy which has not turned passive (rare according to the mass function), or in fact a lower mass but with a 
boosted SSFR due to some special event (like mergers).
In such a case, LIR becomes correlated with the cloud properties because LIR selected sample
get increasingly dominated by galaxies above the main sequence at high LIR.
The global shift to higher LIR in the relation between SED shape and LIR 
for z=1--2 galaxies (Figure~\ref{fg:nuLnu8/LIR rescale curve}) is in the context
of the toy model's Equation~\ref{eq:SFR/m vs. rho} attributable to the increase of 
gas fraction with redshift for galaxies both on and above the
main sequence.

How do the two correlations SED--LIR and SED--$\Delta$SSFR$_{MS}$ fit together? A sloped main sequence produces this naturally:
Over a limited mass range, a selection by LIR (-1 slope in SSFR(M$_*$)) will then produce a similar selection to one by $\Delta$SSFR$_{MS}$ from a sloped main sequence.
As stated before, in the SSFR versus mass diagram we associate the slope of constant \nuLnu(8)/LIR galaxies  with the slope of the main sequence.
Both slopes are not well determined and it is possible that this association, within the large uncertainties, is incidental.
In practice, we measured $\Delta$SSFR$_{MS}$ from a reference line that we fixed.
The more important aspect is that the reference line we chose scales with redshift like the main sequence.

The toy model, capturing cloud properties into ${\rho_c}$ which is assumed to 
be identical for all main sequence galaxies, is admittedly simplifying. 
Nevertheless, it is able to capture the essence of the link between SEDs and 
SSFR offset from the main sequence,
and to motivate scalings with redshift that are related to gas content. 
Further insight will come from testing its assumptions, since ${\rho_c}$ 
may not uniquely define the kind of clouds in the galaxy and not have a 1:1 
link to SED.
Different cloud structures could produce a similar ${\rho_c}$, but different 
SED. Metallicity and non-linear effects can also play a role, in 
particular the latter when the local filling factor of the molecular gas 
approaches $f \sim 1$ and the clouds start affecting each other.
While being a sensitive tracer to the conditions in the star forming regions, \nuLnu(8)/LIR can be affected by many different physical conditions of PAH and continuum emission, grain size distribution and the detailed chemistry.
Attenuation in the mid-infrared in extreme environments, such as those encountered in local ULIRGs can play a role in these special cases.
Therefore some scatter is to be expected in the correlation between \nuLnu(8)/LIR and the distance from the main sequence, LIR, gas depletion time, and other quantities.

\subsection{Comparison of 24~$\mu$m-based SFR estimates adopting different 
mid- to far-infrared SED calibrations} \label{sec:Discussion calibrations}

In Sections~\ref{sec:by SSFR} and \ref{sec:by LIR} we derived two 
calibrations for the 24~$\mu$m photometry as an SFR indicator at redshift 
$\sim$2.
Both produced similar $\sim$0.4~dex scatter in LIR(24)/LIR 
around z$\sim$2 when fitting our data with the modified templates.
The calibration by SSFR is applicable over
a wide redshift range with implicit redshift dependency, whereas the one by LIR is redshift specific.
While we have presented evidence for a smooth variation of \nuLnu(8)/LIR with SSFR 
offset above the main sequence which we consider to be the correct physical 
picture, one can ask about the practical viability of other approaches. 

\begin{figure}[t]
\begin{center}
 \includegraphics[width=1.1\columnwidth]{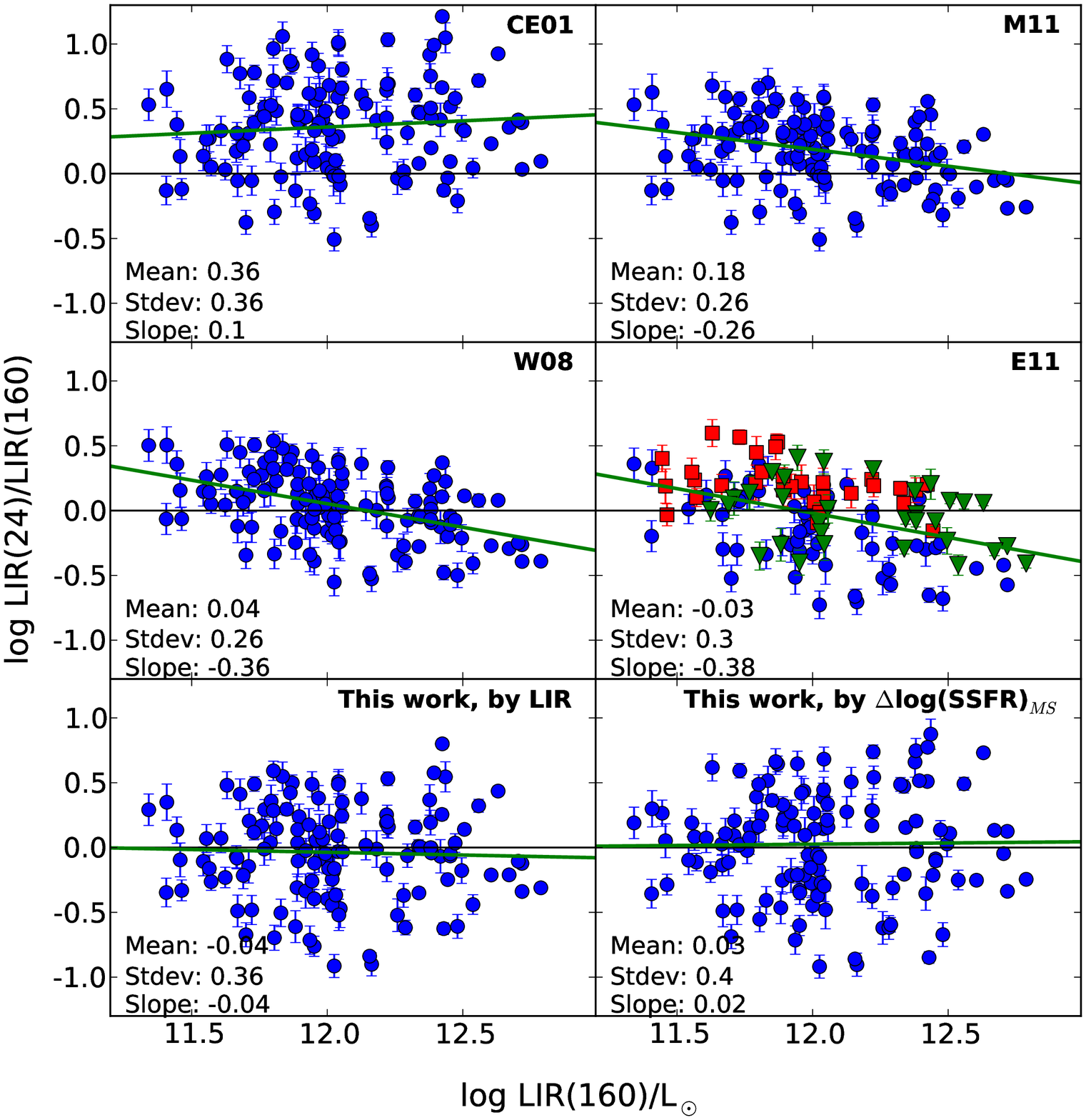}
 \caption{Comparison between various template sets and conversion methods to derive LIR from a single 24 $\mu$m flux measurement. The methods correspond to \citet[][CE01]{CE01}, \citet[][M11]{Murphy11}, \citet[][W08]{Wuyts08}, \citet[][E11]{Elbaz11} and the bottom panels correspond to the calibrations from Section~\ref{sec:by LIR} (left) and Section~\ref{sec:by SSFR} (right). The markers and colors used in E11 correspond to the assumed template: blue circles for main sequence, red squares for starburst, green triangles for a combination of both.
 The green line in each panel is a simple linear trend.}
 %The mean, standard-deviation and the slope of the trend are indicated in each panel.}
 \label{fg:Compare_Nordon_Murphy_Wuyts_Elbaz}
\end{center}
\end{figure}

In the following, we discuss systematic deviations as well as 
the scatter of 
derived LIR(24)/LIR for our z$\sim$2 data, applying our relations as well 
as ones from the literature. We assume only 24~$\mu$m flux, redshift, and 
stellar mass are known. The latter is used only for methods invoking
the SSFR. In the summarizing Figure~\ref{fg:Compare_Nordon_Murphy_Wuyts_Elbaz}, 
we look at the ratio of total infrared luminosity as derived 
from MIPS 24 fluxes LIR(24) to the PACs-based LIR, for the 
1.5$<$z$<$2.5 SFGs in our sample. We 
compare the methods of CE01, \citet[][M11 hereafter]{Murphy11}, 
\citet[][W08 hereafter]{Wuyts08}, \citet[][E11 hereafter]{Elbaz11}, and our 
calibrations by LIR (Equation~\ref{eq:nuLnu8/LIR vs LIR}) and by 
$\Delta$log(SSFR)$_{\rm MS}$ (Equation~\ref{eq:nuLnu8/LIR vs SSFRoffset}).
Note that the x-axis in this figure is the LIR as derived by PACS 
(labeled LIR(160) for clarity) and not 
LIR(24).
Plotting versus LIR(24) would tend to produce positive slopes even for a perfect calibration.
This is because in the presence of noise and a steeply declining luminosity function,
the highest LIR(24) will tend to be dominated by up-scatter of lower luminosities.
In each panel 
we show the results for the individual sources in our sample, the mean offset
of LIR(24)/LIR and the standard-deviation around it, and finally the slope of a 
linear fit that is also overplotted, highlighting any systematic biases.

The intrinsic scatter in \nuLnu(8)/LIR within the galaxy populations seems 
to be significant even after binning by $\Delta$SSFR$_{\rm MS}$ 
(Figure~\ref{fg:nuLnuLir fit by SSFR}), as well as by LIR 
(Figure~\ref{fg:nuLnuLir fit by LIR}), see also 
Figure~\ref{fg:Spearman corr test z2}.
We estimate this scatter to be about $\sim$0.2~dex, 1$\sigma$, consistent 
with findings by \citep{Elbaz11}.
For the purpose of using rest frame 8 $\mu$m for the estimation of LIR, this 
intrinsic scatter sets basic limits to the achievable accuracy of any of the 
conversions. An indirect effect of this scatter arises in calibrations
for which \nuLnu(8)/LIR varies steadily with LIR or with main sequence 
offset, similar to our calibrations and CE01 
(where ${\rm LIR}_{CE01} \propto {\nu}L{_\nu}(8)^{1.5}$). The scatter in the
derived  LIR(24) will then be nonlinearly amplified. For example, 
LIR(24) for a source with a higher than average \nuLnu(8)/LIR (for its LIR)
will not only be overestimated proportional to \nuLnu(8), but in addition 
because a template with lower \nuLnu(8)/LIR  will be applied.
Finally, the scatter will also blur any systematic deficiencies in the 
conversion methods themselves.

The CE01 calibration (Figure~\ref{fg:Compare_Nordon_Murphy_Wuyts_Elbaz} top left)
produces a clear systematic offset for the reasons discussed throughout this
paper. There is no clear trend with luminosity, and the $\sim$0.36 dex scatter
is relatively high because of the nonlinear `amplification'. 

\citet{Murphy11}, on the basis of mostly the data of \citet{Murphy09}, 
suggested an empirical correction to the LIR as estimated when using the 
original CE01 template library. This correction assumes that any
LIR$<10^{12}$~L$_\odot$ as obtained when using the CE01 templates are 
accurate, but applies a luminosity dependent correction when the fit 
results in 
LIR$>10^{12}$~L$_\odot$. This method does not change the templates 
themselves, but instead is applied to the LIR, after it is derived using 
the original CE01 library. This M11 correction by a power 0.6 cancels out 
most of the non-linearity built into the CE01 library, hence also
reduces the associated amplification of scatter.
The calibration of M11 effectively removes the excess seen in the CE01 LIR(24)
for the  LIR$\sim 10^{12.5}$~L$_\odot$ where it was constrained, but the 
luminosity of galaxies with LIR$\lesssim 10^{12}$~L$_\odot$ still tends to 
be overestimated. It thus introduces a trend with the real LIR
(Figure~\ref{fg:Compare_Nordon_Murphy_Wuyts_Elbaz} top right). Note that 
the lower luminosity galaxies used by M11 to constrain their correction are
at lower redshifts, plausibly explaining why systematic offsets remain for
low LIR$\lesssim 10^{12}$~L$_\odot$ galaxies at z$\sim$2.

\citet{Wuyts11a} suggested a unique luminosity-independent conversion 
from 24~$\mu$m to LIR. The conversion is based on a template from 
\citet{Wuyts08}.
This template has log(\nuLnu(8)/LIR)=-0.92 with the MIPS24 filter at z=2. 
This is also the median value of the GOODS-S sample used in this work.
This correction produces for our sample a very good overall correction
(small mean LIR(24) excess),   
but with a clear overall trend with LIR. The luminosities derived by this 
conversion method are overestimated at low LIR and 
underestimated at high LIR.
 (Figure~\ref{fg:Compare_Nordon_Murphy_Wuyts_Elbaz} middle left).
The scatter is low because intrinsic scatter in \nuLnu(8)/LIR is not 
amplified by the method - errors and real variation are propagated linearly.
Given our results, both such a 
residual trend and a moderate scatter will be a feature of correction
recipes that are adopting only a single template. 
In principle, such a single \nuLnu(8)/LIR conversion is in contradiction to 
the clear LIR or SSFR dependence that we find in this study. In practice,
the accuracy of such a conversion depends on the selection of the sample to
which it is applied, and 
will develop biases when selecting mainly low luminosity (main sequence) 
galaxies or very luminous galaxies (high above the main sequence).
Since it does neither encode main sequence offset nor a redshift dependence,
it can also create redshift-dependent biases. Main sequence normalization and 
flux limit of many surveys scale with redshift in a roughly similar way, 
however. This will reduce such effects in practice.

Figure~\ref{fg:CE01 rescale SSFR} suggests that a replacement of the 
continuous SED calibration with SSFR by two discrete values for sources on 
and sources above the main sequence \citep[the approach taken by ][]{Elbaz11} 
can potentially perform well as a two step approximation to the trend.
The change in \nuLnu(8)/LIR between main 
sequence and sources with $\sim$0.5 dex higher SSFR is similar to our
$\sim$0.4~dex scatter around the calibration.
Such a treatment may be in particular applicable
if comparing samples on the main sequence with merger dominated samples
and would directly correspond to the two modes of star formation seen
in CO based studies of the Kennicutt-Schmitt law \citep{Genzel10,Daddi10b}. 
While we do not find such a corresponding bi-modality in the \nuLnu(8)/LIR relation,
it is beyond the scope of this work
to study to which extent the smooth trend in Figure~\ref{fg:CE01 rescale SSFR}
is due to smooth changes of conditions within individual galaxies versus a 
varying mix of `normal' galaxies and `mergers' in a given SSFR bin.

The middle right panel  of Figure~\ref{fg:Compare_Nordon_Murphy_Wuyts_Elbaz} 
shows the result of applying the E11 method to our sample. We have adopted 
here the main sequence parametrization of E11 (mass independent in SSFR) and 
their approach of separating main sequence and `starbursts' at SSFR a factor 2 
above the main sequence. The two templates in E11 were in practice derived by separating the 
galaxies in values equivalent to \nuLnu(8)/LIR and not strictly in SSFR.
Without prior knowledge on the SFR in the galaxy, the selection 
of the correct template can sometimes be ambiguous, with permitted solutions 
for both templates at a given 24~$\mu$m flux and redshift.
The LIR(24) difference then is $\sim$0.25~dex between the two.
These cases are marked as green triangles in 
Figure~\ref{fg:Compare_Nordon_Murphy_Wuyts_Elbaz} and we adopted the mean LIR 
from the two templates. The application of the E11 method to our sample
removes the bias and shows low scatter, but leaves a residual slope.
In these properties, practical results from this two template SSFR-based 
approach resemble the single W08 template method.
  
The template calibrations derived in this work produce negligible trends and 
biases with LIR and $\Delta$log(SSFR)$_{\rm MS}$, but a scatter which is 
1.2--1.5 times larger than the other methods
(Figure~\ref{fg:Compare_Nordon_Murphy_Wuyts_Elbaz} bottom). Much of the 
increase in scatter 
is related to the non-linearity in the conversion from 8~$\mu$m flux to LIR.
This non-linearity in CE01, that also prevails in our calibrations and is 
represented by the strong negative slopes seen in Figure~\ref{fg:nuLnu8/LIR 
rescale curve}, means that a scatter in \nuLnu(8) is super-linearly enhanced 
by a power of $\sim$1.5 when converting to LIR.
With respect to a single template, the improvement by a more detailed 
\nuLnu(8)/LIR versus LIR relation removes bias, but is too 
small to compensate for scatter in \nuLnu(8) and its non-linear increase in 
the conversion to LIR.
When using the template calibrations by $\Delta$log(SSFR)$_{\rm MS}$ (Figure~\ref{fg:CE01 rescale SSFR}, Equation~\ref{eq:nuLnu8/LIR vs SSFRoffset}) the non-linearity is even higher, with a power of $\sim$2.
Deviations in \nuLnu(8) can be due to the photometric errors and due to the intrinsic scatter in the population.
In our sample both have a significant contribution to the overall scatter.
When requiring knowledge of the masses (as when using $\Delta$log(SSFR)$_{\rm MS}$), $\sim$0.2 dex errors in the estimated masses are an additional source
of variation.
Our  $\Delta$log(SSFR)$_{\rm MS}$ calibration introduces a higher 
degree of non-linearity, that increases the other sources of variation by 
larger factors than for other methods. This makes the findings from 
Sect.~\ref{sec:iso nuLnu8/LIR} of best correlations with $\Delta\log({\rm SSFR})_{\rm MS}$, consistent with the larger scatter in LIR(24)/LIR reported here.

Perhaps surprisingly, for 24~$\mu$m-based LIR derivations a simplified linear conversion 
will often produce less scatter in the resulting luminosities than when 
using LIR or $\Delta$log(SSFR)$_{\rm MS}$ dependent templates, even though 
the latter methods better describe the true relation and remove bias.
The difference between the methods and in particular between linear and non linear methods will vary depending on the sample selection and the various sources of noise (photometry, redshifts, masses).
Any attempt of correcting for the sloped trends seen in Figure~\ref{fg:Compare_Nordon_Murphy_Wuyts_Elbaz}, will inevitably include a non-linear scaling that will trade reduced bias for increased scatter.

%%%%%%%%%%%%%%%%%%%%%%%%%%%%%%%%%%%%
\section{Conclusions} \label{sec:conclusions}
%%%%%%%%%%%%%%%%%%%%%%%%%%%%%%%%%%%%
We have obtained deep 70, 100 and 160 $\mu$m {\it Herschel}-PACS photometric 
maps of the GOODS fields as part of the PEP project.
Using this data we study the relation between 8~$\mu$m rest-frame emission and the total IR luminosity \nuLnu(8)/LIR of z$\sim$1 and z$\sim$2 galaxies.
The deep far-infrared observations allow us for the first time to reliably measure the total infrared luminosities of normal star forming galaxies on the z$\sim$2 main sequence with little need of extrapolation or stacking.
We studied the typical SED shapes by binning the galaxies according to 
their $\Delta$SSFR$_{MS}$ distance from the main sequence, as well as by LIR,
We found the SED shape that best describes the population using a photometric 
mean SED fitting method (Section~\ref{sec:method}).
For a subsample of 16 sources at z$\sim$2 and 9 at z$\sim$1 we tested the results from the photometric fit against stacked mid-infrared spectra and found a 
good match between the corrected templates and the spectra in the range 
6--12~$\mu$m.
Finally, we examined a sample of X-ray AGNs which we compared to the SFG population.

Our main findings are summarized below:
\begin{itemize}
 \item The excess in 8 $\mu$m emission with respect to LIR for galaxies at z$>$1.5, reported in earlier studies, can be attributed almost entirely to enhanced emission of PAHs with respect to the locally-calibrated templates that were used. This is verified using a combination of LIR from {\it Herschel}-PACS and stacked deep {\it Spitzer}-IRS spectroscopy.
 No continuum component from an obscured AGN is typically required to explain the mid-infrared emission of far-infrared bright, star forming galaxies with log(LIR/L$_\odot$)$\gtrsim$11.5.
 A similar effect can be observed with a 16~$\mu$m filter at z$\sim$1 (8 $\mu$m rest-frame).
 
 \item \nuLnu(8)/LIR correlates with both the absolute LIR and the $\Delta$SSFR$_{MS}$ distance from the main sequence of star forming galaxies. The latter gives a tighter correlation without explicit redshift dependence \citep[see also ][]{Elbaz11}. 
 We derive calibrations for both relations which we implement on the CE01 template library as an example.
 24 $\mu$m can then be used to estimate LIR in 0.7$<$z$<$2.5 galaxies using the rescaled templates. The scatter is about a factor of 2.5 in our sample, but with no bias or trends with luminosity.
 Calibrating the templates by LIR, z$\sim$1 and z$\sim$2 galaxies have SEDs similar to local galaxies with log(LIR) lower by 0.5~dex and 0.8~dex respectively.
 
 \item The correlation of the SED shape with the distance from the redshift-dependent main sequence suggests that star forming molecular clouds in main sequence galaxies have similar physical conditions at all redshifts z$<$2.5, with only their number and filling factor changing along the main sequence and between redshifts.
 On the main sequence, log(\nuLnu(8)/LIR)=-0.58 when the~8 $\mu$m flux is measured through the MIPS 24 $\mu$m filter at z=2.
 For higher SSFR above the main sequence, \nuLnu(8)/LIR decreases with increasing $\Delta$SSFR$_{MS}$.
 This can be interpreted as more compact star formation, as suggested by \citet{Elbaz11}, in a way 
 consistent with scaling trends observed for other interstellar medium emissions. 
 
 \item The majority of X-ray AGNs in our sample have mid-infrared emission which is completely dominated by the star formation related emission (mainly PAHs).
 This is also seen in a stacked IRS spectrum of 4 X-ray AGNs. Only another, the brightest AGN, shows a significant mid-infrared excess (factor $\sim$3.6) both in photometry and the IRS spectrum, with respect to the (corrected) template of a star forming galaxy.
 We derive a condition on the AGN luminosity and LIR in order for the AGN to contribute an even or greater 8~$\mu$m emission than the star formation.
 The required AGN luminosities for massive z$\sim$2 main sequence galaxies 
 with log(LIR/L$_\odot$)$\sim$12 are high (${\rm L}_{\rm AGN} \gtrsim 10^{45.8}$~erg~s$^{-1}$ ), which makes galaxies with a true 8~$\mu$m excess due to the AGN component rare.
 The fraction of AGNs dominating the 8 $\mu$m emission is strongly dependent on the selection and is likely to increase for log(LIR/L$_\odot$)$\lesssim$11.3.

\end{itemize}

\acknowledgements
PACS has been developed by a
consortium of institutes led by MPE
(Germany) and including UVIE (Austria); KUL, CSL, IMEC (Belgium); CEA,
OAMP (France); MPIA (Germany); IFSI, OAP/OAT, OAA/CAISMI, LENS, SISSA
(Italy); IAC (Spain). This development has been supported by the funding
agencies BMVIT (Austria), ESA-PRODEX (Belgium), CEA/CNES (France),
DLR (Germany), ASI (Italy), and CICYT/MCYT (Spain).

%%%%%%%%%%%%%%%%%%%%%%%%%%%%%%%%%%%%%%%%%%%%
\bibliographystyle{apj}
\bibliography{bibli}
%%%%%%%%%%%%%%%%%%%%%%%%%%%%%%%%%%%%%%%%%%%%%
% latex x bibtex x latex x latex x

%%%%%%%%%%%%%%%%%%%%%%%%%%%%%%%%%%%%%%%%%%%%%%
\appendix
%%%%%%%%%%%%%%%%%%%%%%%%%%%%%%%%%%%%%%%%%%%%%%

%%%%%%%%%%%%%%%%%%%%%%%%%%%%%%%%%%%%%%%%%%%%%%%%%%
\section{A. Determining LIR} \label{app:determin LIR}

\begin{figure}[t]
\begin{center}
  \includegraphics[width=0.7\textwidth]{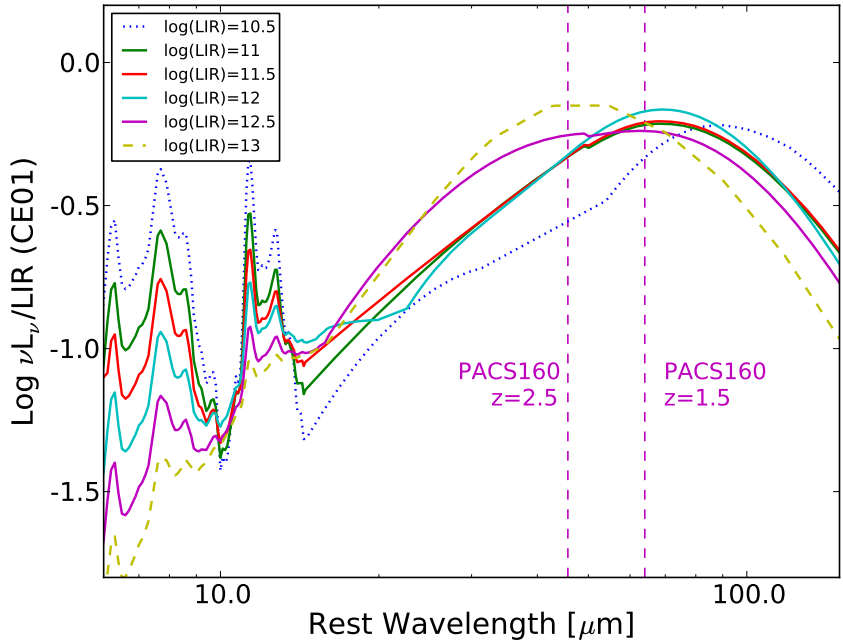} 
  \caption{\nuLnu/LIR for various templates of relevant luminosities from the CE01 SED library.
   Vertical dashed lines indicate the rest-frame wavelengths observed at 160~$\mu$m for z=1.5 and 2.5.}
  \label{fg:nuLnuLirLib}
\end{center}
\end{figure}

The first and most important piece of information we need for each galaxy is 
its 8-1000~$\mu$m LIR. As discussed in Section~\ref{sec:Excess}, the observed 
24~$\mu$m flux is problematic as an LIR estimator for z$\sim$2 galaxies and 
its deviation from the local templates is one of the main issues we wish to 
investigate. We therefore exclude it from the LIR determination which we base
exclusively on the energetically dominant rest-frame far-infrared data. This 
is done also to avoid subtle bias by which, depending on the specific 
adopted weighting scheme, 24~$\mu$m fluxes might affect the LIR from 
template fits to combined mid- and far-infrared data. Only a small 
number of very luminous galaxies are detected 
at 70~$\mu$m and only GOODS-S was observed at this wavelength with PACS, 
which leaves us with PACS 160~$\mu$m and possibly PACS 100~$\mu$m to 
constrain both shape and scale of the SEDs.

Our basic strategy is to derive LIR from fitting CE01 templates to PACS
photometry points close to rest frame 60~$\mu$m.
For reasons described below, at z$\sim$2 we use only the
observed 160~$\mu$m and fit only the scale, maintaining 
the CE01 LIR--SED shape relation. At z$\sim$1 we use both 100 and 160~$\mu$m
and fit both SED shape and scale. We sometimes designate the PACS-based LIR at 
z$\sim$ 2 as LIR(160) for a clear distinction from a MIPS-based LIR(24), but 
refer to 8-1000~$\mu$m rest wavelength in all cases. 

For the higher redshift z$\sim$2 sources, detection rates are limited in the
100 and 70~$\mu$m bands compared to 160 $\mu$m for most luminosity bins.
Selecting only the sources with two far-infrared points will bias our sample 
to the galaxies with bluer far-infrared colors and limit the ability to use 
stacking to compensate for the non-detections.

Fortunately, with {\it Herschel}-PACS we are measuring LIR close to the 
far-infrared peak if using PACS 160~$\mu$m at z$\sim$2.
The determination of LIR is thus much less sensitive to the exact shape of 
the assumed SED.
The flux density at rest frame wavelength of 60 $\mu$m is a particularly good 
measurement for a monochromatic luminosity determination as we shall explain 
below.

\citet{Elbaz10} demonstrated that for redshifts $1<z<3$, fitting the \citet{CE01} SED library to a single PACS 160~$\mu$m measurement results in an excellent agreement with the LIR measured from combined {\it Herschel} PACS+SPIRE photometry (sampling the full far-infrared SED peak).
The robustness of LIR determined from monochromatic 160~$\mu$m luminosity can be understood by looking at the templates normalized to their total IR luminosity. 
The conversion factor from the observed flux F$_{\nu}(\lambda)$ to the total IR luminosity is \nuLnu$(\lambda)$/LIR.
Several templates from the CE01 SED library are plotted in Figure~\ref{fg:nuLnuLirLib} in \nuLnu$(\lambda)$/LIR scale.
The templates for the relevant luminosities of LIR=10$^{11}$--10$^{13}$~\Lsun\ cross each other between 50-60~$\mu$m, where the 160~$\mu$m filter samples the SED at 1.5$<$z$<$2.5 (dashed vertical lines).
Of course, at low enough luminosities such as log(LIR/L$_\odot$)=10.5 ($\sim$3.5~M$_\odot$~yr$^{-1}$) the template deviates for the typical \nuLnu/LIR at 60 $\mu$m.
However, our typical luminosities are more than an order of magnitude higher than that and such cold SEDs will not be common.

Nearly identical \nuLnu$(\lambda)$/LIR values for the different templates at a given wavelength mean that for the redshifts at which the filter is centered on this wavelength, LIR can be measured regardless of the specific template that is selected.
Possible systematic errors due to selecting a template different from the optimal one are very small and can be visually seen in Figure~\ref{fg:nuLnuLirLib} as the distance between the SED lines (up to 0.1~dex around 60 $\mu$m).
This fortunate coincidence around 60~$\mu$m will gradually worsen if the dust temperatures get significantly lower than the local LIR-T relation implemented in the CE01 library.
\citet{Hwang10} studied deep {\it Herschel} data from the far-infrared to sub-millimeter and report that this is not the case and this result agrees with the accuracy of LIR measured from 160 $\mu$m as found by \citet{Elbaz10}. Further support
is given to this by the deeper data of \citet{Elbaz11}.
We therefore measure LIR for z$\sim$2 galaxies by fitting CE01 templates to 160~$\mu$m fluxes alone, even in cases where shorter wavelength fluxes are available.

The reason not to use PACS 70 and 100 $\mu$m even when available is twofold:
One is that the fairly low detection rates at these wavelengths. If we use the 
added data points per galaxy to relax the luminosity--template relation in the library
it means that we will be treating the part of the population which is detected at 
100 $\mu$m differently than the non-detections.
The other is that the z$\sim$2 70 and 100 $\mu$m bands observe 
$\sim$25 and 35 $\mu$m rest, far from the SED peak and a region of the SED 
which is not part of the modified blackbody emission and for which 
the \nuLnu$(\lambda)$/LIR values can vary significantly between templates 
(Figure~\ref{fg:nuLnuLirLib}).
Using this SED region to constrain LIR means that we allow a measurement 
that represents a small fraction of the total luminosity to determine LIR. 
This can lead to systematic errors if the selected template is inaccurate
and even larger errors if we do not thaw the luminosity--template relation in the library.
The combination of these two effects makes the potential biases and 
systematic errors outweigh the gain in added constraints to the LIR 
determination.

\begin{figure}[t]
\begin{center}
  \includegraphics[width=0.7\textwidth]{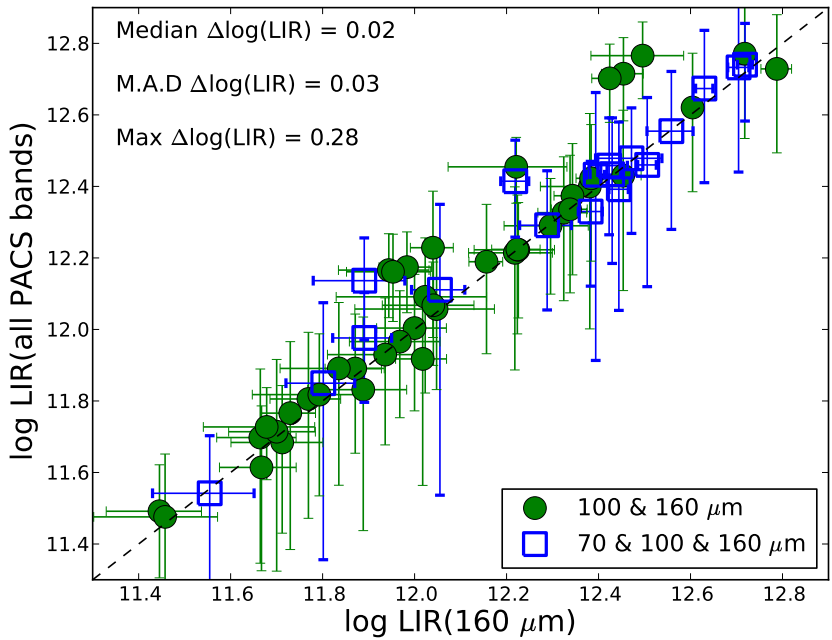} 
  \caption{Comparison of total infrared luminosity LIR(all PACS bands) derived from all available PACS bands (at least two detections), versus total infrared luminosity LIR(160) from 160 $\mu$m alone. LIR(all PACS bands) is determined by fitting all available PACS bands with free scale and template shapes in the CE01 library.
  LIR(160), used throughout the paper, is determined by fitting 160 $\mu$m alone, maintaining the original luminosity--template relation in the library.
  We select all 1.5$<$z$<$2.5 galaxies in our sample that have two or three PACS fluxes (45 and 15 out of 128).
  The statistics for $\Delta\log({\rm LIR}) = \log({\rm LIR}({\rm all\,PACS\,bands}))-\log({\rm LIR}(160))$ are given in the figure. Errorbars represent formal random errors only, propagated from the photometric errors.}
  \label{fg:LIR160_vs_LIRPACS_compare}
\end{center}
\end{figure}

In any case, the difference when including or excluding the shorter wavelengths is quite small as we demonstrate in Figure~\ref{fg:LIR160_vs_LIRPACS_compare}.
In our z$\sim$2 sample, considering only sources that have 160 $\mu$m and at least one other PACS band detections (half of our full z$\sim$2 sample), we fit all available PACS bands with CE01 templates (free template and scales, $\chi^2$ minimization) and compare with the LIR derived by a monochromatic 160~$\mu$m fit. The median difference and median absolute deviation (MAD) between the two methods are 0.02 and 0.03 dex and the maximum difference is 0.28 dex.
The added (and possibly biased) constraint provided by the 100 and 70 $\mu$m points is used for selecting a slightly different template, that still results in nearly the same LIR which is determined by the 160 $\mu$m flux almost independently (see Figure~\ref{fg:nuLnuLirLib}). Only 5 of 60 galaxies deviate in their monochromatic LIR(160) from the multi-band LIR(PACS) by more than 1-$\sigma$, the largest of which is 1.5 $\sigma$.
One should keep in mind that due to the detection limits, many galaxies with real 100/160 $\mu$m flux ratios that are lower than our sample values will simply not be detected in 100 $\mu$m and are not represented in this comparison.

Concerning our lower redshift z$\sim$1 bin, the far-infrared rest-frame 
wavelengths scanned by the three PACS photometers band for $0.7<z<1.3$ 
spans across 20--95~$\mu$m, overlapping in large part the range covered for 
$1.5<z<2.5$ (20--65~$\mu$m).
At these lower redshifts, the PACS 160~$\mu$m filter has moved farther away 
from the rest frame 60~$\mu$m and PACS 100 $\mu$m filter has moved closer.
The lowest luminosities in the lower redshift sample are of 
log(LIR/L$_\odot$)$\approx$10.8, which may include colder SEDs than those
for which the 60~$\mu$m derived LIR is ideal.
By coincidence, for the lower luminosities of the z$\sim$1 sample (log(LIR/L$_\odot$)$\lesssim$12), the templates cross at rest 80--90 $\mu$m (Figure~\ref{fg:nuLnuLirLib}), which is again the rest wavelength of the 160 $\mu$m filter.
At these redshifts 100 $\mu$m has a much higher detection rate (100\% of the 160 $\mu$m sources in most bins we use) and probes a much more useful rest-frame wavelength ($\sim$50 $\mu$m) than it does at z$\sim$2.
The arguments for using only 160 $\mu$m mentioned above do not apply to this sample.
For LIR determination at z$\sim$1, we fit the best combination of CE01 template shape and scale (as two free parameters) to the 100 and 160 $\mu$m points simultaneously. For a few cases at the lowest luminosities where only 160~$\mu$m is available and then a monochromatic fit to LIR is used.

%%%%%%%%%%%%%%%%%%%%%%%%%%%%%%%%%%%%%%%%%%%%%%%%%%%%%%%%%%%%%
\section{B. Stacking analysis} \label{app:stacking}

  \subsection{Measuring the stacked flux}

Sources which are undetected in 70 or 100 $\mu$m (but detected at 160~$\mu$m) are stacked to produce a mean photometric point representing this population.
Stacking is performed into a residual map from which all individually detected sources have been removed. This is done in order to minimize the contamination from much brighter nearby sources.
The viable depth of stacks is limited by the density of neighboring sources which are brighter than the typical sources in the stack.
In this work we stack into the 70 and 100 $\mu$m images on the position of MIPS 24~$\mu$m sources, which are detected at 160~$\mu$m.
This means that the brightness of our stacked sources is likely just below the detection threshold, or they will not have been detected at 160~$\mu$m. By removing the individual detections, our sources of interest are among the brightest sources that are still left in the image.

The background level for each stacked image is estimated from the position of the peak in the pixel value distribution $P(D)$. This method is more reliable than measuring the background from the stacked image itself.
Due to selection effects, prior positions will tend to be at a minimum distance from the neighboring sources, thus the priors tend to be positioned at local minima in the map, which is reflected as a depressed background-level at the center of the stacked image.
Background estimated from the edges of the stacked image will tend to be higher than the background at the position of the stacked source.
Subtracting the background according to the $P(D)$ peak and extracting the flux using a limited-radius point-spread function avoids this issue.

%The reason the $P(D)$ peak represents the background level is the following: instrumental noise is distributed around the diffuse background flux, while individual sources in the image add only positive flux to the image. The pixel flux distribution from many point sources (and only sources) will have a peak at some positive value that depends on the number counts and the pixel size, but will drop to zero at zero flux. Thus the real sources in the map affect the wings of the P(D) distribution, but have little effect on the peak. The background level in the PACS images is arbitrary and any diffuse background or highly confused sources that mimic a diffuse background are removed by the high-pass filter, which was applied to the time-lines of the detector pixels during reduction.

\subsection{Stacked \nuLnu/LIR}
The value of interest for the stack is \nuLnu$(\lambda_0)$/LIR, at a certain rest frame wavelength $\lambda_0$. When converting from stacked flux to \nuLnu$(\lambda_0)$/LIR one must account for the different redshifts of the stacked sources (affecting both the luminosity-distance and the rest wavelength), as well as for the different (known) individual LIR.
We use the following weighted mean to get the stack mean \nuLnu/LIR:
\begin{equation}
\lambda_0 \equiv \lambda_{filter}/(1+\bar{z})
\end{equation}
\begin{equation}
{\nu}L_{\nu}(\lambda_0)/LIR \bigr|_{stack} = N_{stack}\cdot F_{\nu}(\lambda_{filt}) \left[ \displaystyle\sum_{i=1}^{N_{stack}}{\frac{\lambda_{filt} \left( \frac{1+z_i}{1+\bar{z}} \right)^\alpha LIR_i}{c 4{\pi}D_i^{2}} } \right]^{-1}
\label{eq:weightednuLnu}
\end{equation}
where $\lambda_{filt}$ is the filter central wavelength, $\lambda_0$ is the adopted rest-frame wavelength for the stacked flux. $\bar{z}$ and $z_{i}$ are the mean and source-$i$ redshifts, $N_{stack}$ is the number of sources in the stack, $F_\nu$ the stacked observed flux and $D_i$ the luminosity-distance for each source.
$\alpha$ is the approximated slope of the SED (${\nu}L_{\nu} \propto \lambda^{\alpha}$) at $\lambda_0$, which is used to correct the fluxes to a common $\lambda_0$.
The exact value of $\alpha$ makes only small difference in the limited redshift intervals we use - we assume $\alpha \approx 1.5$ close to the slope of LIRG templates in CE01, which is justified posteriorly.

\subsection{Uncertainty on the stacked flux}
There are two kinds of uncertainties regarding the stacked flux as an estimator to the mean flux of the population: one is photometric, i.e. the significance of the detection of a source in the stacked image. The other is on the stacked flux as the estimator to the true mean flux of the stacked population as derived from a finite size sample. The latter depends on the flux distribution within the individual sources in the stack.

In order to estimate the photometric uncertainty $\sigma_{phot}$ we create a large number of additional stacks, each using $N$ random positions in the image, where $N$ is the number of stacked sources in the real stack.
Each random stack is treated in the same way the real stack and a source flux is measured. The standard deviation in the fluxes measured from the random stacks is taken as the absolute flux uncertainty.
This determines the stacked-source detection significance, i.e. the probability to produce such a flux measurement from a completely meaningless stack.
For stacks without a detected source (flux$< 3 \sigma_{phot}$) a flux of 3$\sigma_{phot}$ is taken as the upper limit.

The uncertainty on the stacked flux as an estimator to the {\it mean} source flux of the stacked population is estimated using a bootstrap method:
From the list of priors in the stack a new list, identical in size is created by random re-sampling with replacements.
We perform a large number of these re-samplings and produce a distribution of measured fluxes.
A 68\% confidence interval is calculated from this distribution.
The bootstrap error on the mean stacked flux of large stacks already includes photometric flux errors. However, on a small stacked sample they might be underestimated.
Therefore the larger of the two is taken as the final uncertainty.

%%%%%%%%%%%%%%%%%%%%%%%%%%%%%%%%%%%%%%%%%%%%%%%%%%%%%
\section{C. The mean-SED redshift-scan method: $\chi^2$ minimization} \label{app:chisqr_minimization}

When fitting an SED template to the collection of photometric points from the subsample and applying a $\chi^2$ minimization, one must take into account that these measurements are not repeated measurements of the same source. 
The measurements scatter not only due to the errors related to the photometry of each measurement $\sigma_{phot}$, but also due to the intrinsic scatter in the population $\sigma_{pop}$.
Neglecting to account for this will result in a biased fit which is dominated by the few points with the best signal to noise; these would be the galaxies with the bluer far-infrared colors that are better detected at 70 and 100 $\mu$m than the redder ones.
$\sigma_{pop}$ the intrinsic scatter in \nuLnu$(\lambda_0)$/LIR of the population must be estimated and included in the standard deviation for each data point when calculating the $\chi^2$ to be minimized:
\begin{equation}
 \sigma^{2}_{{\nu}L_\nu/\rm{LIR}} = \sigma^{2}_{phot} + \sigma^{2}_{pop}
\end{equation}

The inclusion of $\sigma_{pop}$ has an important implication: for cases in which $\sigma_{pop} \gtrsim \sigma_{phot}$, measurements of different sources in a given filter have nearly the same weight regardless of the photometric error. We must emphasize again that this is the appropriate weighting when doing a repeated sampling of a population rather then repeated measurements of the same object.

When estimating the intrinsic \nuLnu/LIR scatter in a population of galaxies, two things need to be taken into account: one is that the photometric points from different sources are for different rest frame wavelengths and need to be k-corrected to a common wavelength.
The other is that for most cases, a significant fraction of the sources are not detected individually and are included as a stacked mean instead.

When the significant majority of the sources (more than 80\%) are detected by the given filter, we can ignore the small fraction of undetected sources and estimate the scatter from the detections only.
For this purpose we correct all \nuLnu/LIR of the same filter to a common rest-frame wavelength (assuming a ${\nu}L_{\nu} \propto \lambda^{1.5}$ approximated slope) and calculate the standard deviation in \nuLnu/LIR.
When binning the galaxies by deviation from the main sequence, $\sigma_{pop}$ is also estimated from the detections only:
In a selection by constant SSFR, the fainter, non-detected sources tend to be those of galaxies of lower absolute luminosity, not necessarily of lower \nuLnu/LIR.
Therefore, detections and non-detections are mixed in \nuLnu/LIR of the far-infrared filters.
While a weak trend with luminosity does exist, it is a secondary effect for $\sigma_{pop}$ estimation.

When selecting by LIR and a significant fraction (over 20\%) of the population is individually undetected in the given filter, the scatter is unresolved in its lower part and the weighted mean of the distribution is yet to be determined.
In  order to estimate the standard deviation of the full population, we approximate the distribution to be normal. In this case we can use the fact that in a narrow LIR bin the detected and non-detected source populations are fairly well separated by the detection threshold. 
For the stacked sources we only know their mean \nuLnu($\lambda_0$)/LIR and we calculate the same value for the detections using equation~\ref{eq:weightednuLnu}.
We then find the standard deviation of a normal distribution, that when split into two sub-populations at a value $x_c$, the sizes of the two sub-populations ($x<x_c$ and $x>x_c$) have the same ratio as $N_{detect}/N_{stack}$ and the difference between their means
match the difference in the means of the detections and non-detections ($\mu_{detect}-\mu_{stack}$).
This problem has a simple and easy numerical solution.

% The $x_c$ parameter for a normal distribution $f(x)$ is determined from the inverse of the normal cumulative distribution function:
% \begin{equation}
%  \frac{N_{stack}}{N_{stack}+N_{detect}} = \int_{-\infty}^{x_c} f(x) \mathrm{d}x
% \end{equation}
% %Since we used a normal distribution with $\sigma = 1$
% This yields $x_c$ in units of the standard deviation (STDEV) of $f(x)$.
% The difference in the means of the two subpopulations, created by splitting a normal distribution at $x_c$ is:
% \begin{equation}
%  \Delta \mu_{normal} = \frac{ \int_{x_c}^{\infty} xf(x) \mathrm{d}x }{ \int_{x_c}^{\infty} f(x) \mathrm{d}x } - \frac{ \int_{-\infty}^{x_c} xf(x) \mathrm{d}x }{ \int_{-\infty}^{x_c} f(x) \mathrm{d}x }
%  = (N_{stack}+N_{detect}) \left[ \frac{\int_{x_c}^{\infty} xf(x) \mathrm{d}x}{N_{detect}} - \frac{\int_{-\infty}^{x_c} xf(x) \mathrm{d}x}{N_{stack}} \right]
% \end{equation}
% Which is easy to integrate numerically.
% Due to the way we defined $x_c$ the difference between the means $\Delta \mu_{normal}$ is also expressed in units of STDEV.
% >From here, to get $\sigma_{pop}$ the relations is:
% \begin{equation}
%  \sigma_{pop} = \frac{ \mu_{detect}-\mu_{stack} }{ \Delta \mu_{normal} }
% \end{equation}

%%%%%%%%%%%%%%%%%%%%%%%%%%%%%%%%%%%%%%%%%%%%%%%%%%5
\section{D. Calibrating a template library by $\Delta$log(SSFR)$_{\rm MS}$} \label{app:calibrate library}

In this appendix we describe how to apply the calibration derived in Section~\ref{sec:by SSFR} to a template library, expressed as a function of \nuLnu(8)/LIR or LIR, which can then be fitted to a galaxy with a known mass and a redshift.
The additional implicit information we need is the redshift-dependent main sequence SFR($M_*,z$)$_{\rm MS}$, that was defined in Section~\ref{sec:method} and is based on the calibration of \citet{Rodighiero10}.
The distance from the main sequence is then $\Delta\log({\rm SSFR})_{\rm MS} = \log({\rm SFR}/M_*)-\log({\rm SFR}(M_*,z)_{\rm MS}/M_*)$.
We can use Equation \ref{eq:nuLnu8/LIR vs SSFRoffset} or \ref{eq:LCE01 vs SSFRoffset} to eliminate $\Delta\log({\rm SSFR})_{\rm MS}$ and find a direct association between a template defining property (\nuLnu(8)/LIR or $\Lambda_{\rm CE01}$) and the LIR it should be scaled to.
We will proceed with the specific example of the CE01 library and define $\chi$ as the ratio LIR/SFR (Equation~\ref{eq:LIR2SFR}):
\begin{equation}
 \log \left( \frac{LIR}{L_\odot} \cdot \chi \right) = \frac{ \log(\Lambda_{\rm CE01}/L_\odot)-10.58 }{ 1.44 } + \log \left( \frac{SFR(M_*,z)_{\rm MS}}{M_\odot {\rm yr}^{-1}} \right)
\end{equation}
Thus, we scale each of the log($\Lambda_{\rm CE01}$/L$_\odot$)$>$10.38 templates in the original library to the new LIR, simply by applying a factor $LIR/\Lambda_{\rm CE01}$ to the template.
The log($\Lambda_{\rm CE01}$/L$_\odot$)$<$10.38 templates in the library either have \nuLnu(8)/LIR$>$-0.58 or quickly become degenerate in this value.
Our calibration does not deal with these low luminosity templates, that tend to represent galaxies well below the current IR detection limits at the redshifts of interest.

%%%%%%%%%%%%%%%%%%%%%%%%%%%%%%%%%%%%%%%%%%%%%%%%%%%%%%%%%%%%%%%%%%%%%%%%%%%%%%%%%
%%%%%%%%%%%%%%%%%%%%%%%%%%%%%%%%%%%%%%%%%%%%%%%%%%%%%%%%%%%%%%%%%%%%%%%%%%%%%%%%%
%%%%%%%%%%%%%%%%%%%%%%%%%%%%%%%%%%%%%%%%%%%%%%%%%%%%%%%%%%%%%%%%%%%%%%%%%%%%%%%%%

\end{document}